\documentclass[aps,prd,twocolumn,groupedaddress,amsmath,amssymb,nofootinbib]{revtex4-2}

\usepackage{graphicx}
\usepackage{dcolumn}
\usepackage{bm}
\usepackage{hyperref}
\usepackage[capitalize]{cleveref}
\crefname{section}{Sec.}{Secs.}
\Crefname{section}{Section}{Sections}
\usepackage{lipsum}
\usepackage{braket}
\usepackage{xcolor}
\usepackage{bbold}				
\usepackage{mathtools}

\newcommand{\lParen}{(}
\newcommand{\rParen}{)}

\newcommand{\bzeta}{\boldsymbol{\zeta}}
\newcommand{\btheta}{\boldsymbol{\theta}}
\newcommand{\bfeta}{\boldsymbol{\eta}}
\newcommand{\blambda}{\boldsymbol{\lambda}}

\newcommand{\bpi}{\boldsymbol{\pi}}

\newcommand{\bfz}{\bm{0}}

\DeclareMathOperator{\ch}{\text{ch}}
\DeclareMathOperator{\sh}{\text{sh}}
\DeclareMathOperator{\proj}{\mathcal{P}}

\newcommand{\SymGroup}{\mathrm{Sym}}

\newcommand{\mass}{m}

\newcommand{\scale}{\sigma}

\makeatletter
\newcommand{\oset}[3][0ex]{%
  \mathrel{\mathop{#3}\limits^{
    \vbox to#1{\kern-1\ex@
    \hbox{$\scriptstyle#2$}\vss}}}}
\makeatother
\newcommand{\rev}[1]{\oset{\leftarrow}{#1}}

\newcommand{\finfty}{F_\mathrm{min}^{\infty}}

\begin{document}


\title{Numerical results on Quantum Energy Inequalities in
Integrable Models at the Two-Particle level}


\author{Jan Mandrysch}
\email[Email: ]{jan.mandrysch@fau.de}
\affiliation{Department Mathematics, Friedrich-Alexander University Erlangen, Germany}
\affiliation{Institut for Theoretical Physics, Leipzig University, Germany}


\date{\today}

\begin{abstract}
    In this article, we study the impact of self-interaction and multi-particle states on sustaining negative energies in relativistic quantum systems. For physically reasonable models, one usually requires bounds on both magnitude and duration of the accumulation of negative energy, typically given in form of a Quantum Energy Inequality (QEI). Such bounds have applications in semiclassical gravity where they exclude exotic spacetime geometries and imply the formation of singularities.

    The essence of this article is a novel numerical method for determining optimal QEI bounds at the one- or two-particle level, extending previous work focused on the one-particle case and overcoming a new type of technical challenge associated with the two-particle scenario. Our method is tailored for integrable models in quantum field theory constructed via the S-matrix boostrap.

    Applying the method to a representative model, the sinh-Gordon model, we confirm self-interaction as the source of negative energy, with stronger interactions leading to more pronounced negativities. Moreover, we establish the validity of QEIs and the averaged weak energy condition (AWEC) at the one- and two-particle level. Lastly, we identify a constrained one-parameter class of nonminimal stress tensor expressions satisfying QEIs at both levels, with more stringent constraints emerging from the QEI bounds at the two-particle level.
\end{abstract}


\maketitle

\section{Introduction \label{intro}}
In general relativity, Einstein's equations couple the geometry of spacetime to matter which is represented by the so-called stress(-energy) tensor. To select physically reasonable solutions one typically imposes energy conditions. The classical energy conditions---such as the weak energy condition (WEC) or the null energy condition (NEC)---represent pointwise positivity of the energy density and pressure as measured by moving observers. On the geometric side these conditions exclude exotic spacetime configurations like wormholes or warp drives and imply the formation of a singularity provided that the geometry has contracted sufficiently to form a trapped surface \cite{Cur17}.

Quantum matter, though, has challenged this perspective: Negative energy densities appear abundantly in quantum field theory \cite{EGJ65} as for example in the well-tested phenomenon known as Casimir effect. The energy density can even scale to negative infinity at a point; see e.g. \cite[Sec.~2]{Few06}. Thus all classical energy conditions of GR are violated by quantum matter. Alleviating is that weaker alternatives have been developed. On the one side there are the averaged energy conditions (AWEC, ANEC, ...) which require the classical energy conditions to hold ``globally on average'' meaning when integrated along full trajectories in spacetime. On the other side, there are the so-called \emph{quantum energy inequalities} (QEI) which retain the local character of the classical energy conditions but allow for bounded negative energies. 

Let us denote the stress-energy measured along an inertial trajectory $\gamma(\tau)$ with velocity vector $u$ and averaged over a positive test function $g^2$ by
\begin{equation}
    T(\gamma,g^2) = \int d\tau \, u^\mu u^\nu T_{\mu\nu}(\gamma(\tau)) g^2(\tau),
\end{equation}
where $T_{\mu\nu}(x)$ is the stress tensor at spacetime point $x$. Focussing on timelike trajectories $\gamma$, a QEI may take the following form:
\begin{equation}
      T(\gamma,g^2) \geq - c_g \, \mathbb{1}\label{eq:introQEI}
\end{equation}
for a constant $c_g > 0$ which depends only on $g$. If the inequality holds in a suitably large set of states, we say that a state-independent (worldline) QEI holds. However, it is possible that one has to allow for a mild dependence of $c_g$ on the states; in this case we call the QEI state-dependent. Note here that averaged energy conditions can in principle be inferred by studying the limit $g \to 1$. For instance, the AWEC requires that
\begin{equation}
    T(\gamma,1) = \int d\tau \, u^\mu u^\nu T_{\mu\nu}(\gamma(\tau)) \geq 0 \label{eq:awec}
\end{equation}
for all inextendible timelike geodesics $\gamma$ and when the integral is absolutely convergent. It is implied from \eqref{eq:introQEI} provided that $c_g \to 0$ in the limit $g \to 1$.

Both types of weakened energy conditions (QEIs and averaged conditions) are still effectively imposing constraints on exotic spacetimes and allow to state conditions for singularity formations similar as with classical energy conditions; see \cite{KS20} for a review.

QEIs have been proven for many kinds of free QFT models on flat and curved spacetimes and also conformal field theories in 1+1d \cite{For91,FR95,FR97,PF97,Fla97,FE98,FPR98,PF98,PF98a,Vol98,FT99,Few00,Vol00,FV02,Pfe01,FM03,GO03,Few04,GOS04,FH05,Daw06,DF06,SYL+06,FO08,FS08,SV08}. Also averaged energy conditions, here referring to its weakest variants ANEC and achronal ANEC, are known to hold for scalar and vector fields in free QFTs on Minkowski space \cite{Kli91,Fol92a,FR95,Yur95,FW96,FR96,FR03,FH05}
and restricting to 1+1d also for general situations with interaction and curvature \cite{WY91,Ver00,FH05}.

For QEIs, though, their validity in the presence of self-interaction is less clear. There exist general model-independent inequalities which are however weaker than the previously discussed ones \cite{BF09,MPV22} and it is unclear if they are able to impose constraints on the geometric side. 
More is known when specializing to the class of 1+1d integrable models. A QEI in the presence of self-interaction was first established in the Ising model by using an analogy with the free fermionic field \cite{BCF13}. The only other full QEI in this context pertains to a recent result for the sine-Gordon model in the superrenormalizable regime with an adiabatic cutoff \cite{FC22}. QEIs at one-particle level, i.e., where \eqref{eq:introQEI} holds when evaluated in one-particle states, have been obtained for the class of integrable models with one scalar particle type \cite{BC16} and recently generalized to models with bound states, arbitrary particle types, and inner degrees of freedom \cite{BCM23}.

In this article, we present a method to obtain numerical estimates for sharp QEI bounds in one- and two-particle states in the class of integrable models: We will analyse the spectrum of a discretized version of the averaged stress tensor in one- and two-particle states. The lower bound of the spectrum will serve as an approximation for the optimal choice of the constant $c_g$ appearing in \eqref{eq:introQEI}. For the time being, we will limit ourselves to treat a simple representative model, the sinh-Gordon model, which describes a single scalar particle without bound states. The method, however, is general and in principle applies to any integrable model provided that the truncated momentum space correlation functions---also known as form factors---of the stress tensor are known to low order.

While the general strategy is similar to a numerical analysis for the one-particle case \cite{BC15}, the two-particle case has two additional significant challenges. First, going from one- to two-particle states, the numerical dimension of the problem squares so that computations are much more costly. Second, while the to-be-discretized kernel is analytic at the one-particle level, at the two-particle level it has singularities. This needed extra attention and a careful inclusion in the numerical approximation in order to achieve numerical stability.

The numerical results for the sinh-Gordon model indicate that QEIs at one- and two-particle level share many qualitative features but accumulation of negative energies can be significantly higher in magnitude and duration in states with two particles. At the one-particle level, we find agreement with the results obtained in \cite{BC15}. The results at two-particle level are new.

Concluding with the general introduction (\cref{intro}), we will start with a brief but self-contained account on the treatment of observables in integrable models (\cref{integrable}). As part of this section, we will specifically discuss the stress tensor (\cref{ff,nonmin}) and introduce the sinh-Gordon model (\cref{shg}). Next, we introduce the numerical setup. This includes a brief account on the physical input data (\cref{physical}), a general description of the discretization method (\cref{discretization}) as well as a detailed description of the central methodological challenge, the treatment of singularities (\cref{explcomp,explexpr,explexprnlo}). It also includes some further methodological aspects (\cref{numminsol,spectral,plausibility}). Finally, we present our numerical results (\cref{results}) including plots of lower bounds for the stress-tensor at one- and two-particle level with varying model inputs. We conclude on these results in \Cref{conclusion}.

\section{Integrable models, observables and form factors}\label{integrable}

In this section we will briefly review the setup of integrable models via the inverse scattering approach---also referred to as S-matrix bootstrap---and give the description of observables in terms of their form factors. The starting point of the construction consists of fixing the particle content of the model and a scattering function representing the two-to-two particle interactions. Assuming integrability, this suffices to determine the full state space and local field content of the model. While this applies to general models with several particle types and inner degrees of freedom \cite{LS14,AL17,BCM23,Man23}, we will confine ourselves to treat models with a single bosonic scalar degree of freedom and in particular the sinh-Gordon model serving as a representative example.

In our setup, the (two-to-two particle) scattering function can be described by a single complex-valued function $S(\theta)$ conveniently parametrized by the rapidity difference $\theta$ of the incoming (or outgoing) particles. Typical properties in scattering theory like unitarity and crossing symmetry amount to $S$ being analytic in the physical strip $\mathbb{R}+i[0,\pi]$ and satisfying the relations \begin{eqnarray}
    S(\zeta)^{-1} = S(-\zeta) = S(\zeta+i\pi) = \overline{S(\bar\zeta)}, \quad \zeta \in \mathbb{C}.
\end{eqnarray}

The full state space for integrable models is then given by an $S$-symmetrized version of second quantization \cite{LM95}: With one-particle space $\mathcal{H}_1=L^2(\mathbb{R})$ the interacting $k$-particle space is given by
\begin{eqnarray}
    \mathcal{H}_k = L^2_S(\mathbb{R}^k) = \{ \psi \in L^2(\mathbb{R}^k), \psi \text{ S-symmetric} \},
\end{eqnarray}
where $\psi$ is referred to as \emph{S-symmetric} iff for all $\btheta \in\mathbb{R}^k$ and $\tau\in\SymGroup_k$,
\begin{eqnarray}
    \psi(\btheta) = S^\tau(\btheta) \psi(\btheta^\tau) :=  \prod_{\mathclap{\substack{i< j \\\tau(i) > \tau(j)}}} S(\theta_{\tau(i)}-\theta_{\tau(j)}) \psi(\btheta).
\end{eqnarray}
Here, $\btheta^\tau = (\theta_{\tau(1)},\ldots,\theta_{\tau(k)})$ and $\SymGroup_k$ denotes the symmetric group with $k$ elements.
As further notation, we introduce $\proj^{(k)}_S$ which projects from unsymmetrized Fock space onto S-symmetric $k$-particle vectors,
\begin{eqnarray}
    \proj^{(k)}_{S}\psi(\btheta) = \sum_{\tau \in \SymGroup_k} S^\tau(\btheta) \psi(\btheta^\tau),
\end{eqnarray}
Occasionally, we will use $\proj^{(k)}_{S,\btheta}$ to specify on which variables $\proj^{(k)}_S$ acts.

To describe observables, it is customary to introduce improper rapidity eigenstates $\ket{\btheta} \equiv \ket{\theta_1,\ldots,\theta_k}$ which are fixed by orthonormality,
\begin{eqnarray}
    \braket{\btheta | \bfeta} = \proj^{(k)}_{S,\btheta} \delta(\btheta- \bfeta),
\end{eqnarray}
and the eigenrelation,
\begin{eqnarray}\label{eq:eigen}
    P^\mu \ket{\btheta} = p^\mu(\btheta) \ket{\btheta}, \quad  p^\mu(\btheta) := \sum_{j=1}^k p^\mu(\theta_j).
\end{eqnarray}
Here $P^\mu$, $\mu=0,1$, denotes the total energy momentum operator and $p(\theta) = \mass \,( \ch \theta,  \sh \theta)^t$ a single on-shell momentum parametrized by the rapidity $\theta$.

The $k$-particle form factor of an observable $A$ is then given by
\begin{eqnarray}
    F_{k}(\btheta) = \sqrt{k!} \braket{\btheta | A |\Omega},
\end{eqnarray}
where $\ket{\Omega}$ is the vacuum vector. The expectation value of $A$ in a state $\psi$ takes the form
\begin{eqnarray}\label{eq:expval}
    \braket{\psi, A \psi} = \int d\btheta d\bfeta \, \overline{\psi(\btheta)} \braket{\btheta | A |\bfeta} \psi(\bfeta).
\end{eqnarray}

Assuming that $A$ is localized in a finite region, the $F_k$ allow for a meromorphic continuation and satisfy the so-called \emph{form factor equations} \cite{KW78,BFK08,BC13,BC15}\cite[Chap.~3]{Man23}. Assuming for simplicity, that all odd form factors vanish, i.e., $F_{2k+1} = 0$, the relation of \eqref{eq:expval} to the form factors is then given by
\begin{widetext}
\begin{eqnarray}\label{eq:kernel}
    A(\btheta,\bfeta) := \braket{\btheta | A |\bfeta} =  \proj^{(k)}_{S,\btheta} \proj^{(k)}_{S^{-1},\bfeta} \sum_{j=1}^{k} \tfrac{k!}{(j!)^2(k-j)!} F_{2j}(\btheta_{1..j}+i\bfz,\rev{\bfeta}_{1..j}+i\bpi-i\bfz) \delta(\theta_{j+1}-\eta_{j+1}) \ldots \delta(\theta_k-\eta_k),
\end{eqnarray}
\end{widetext}
where $\btheta_{i..j} = (\theta_i,\ldots,\theta_j)$ \footnote{This expression compares with the inversion formula derived in \cite[Prop.~3.5]{BC13} using $\bra{\btheta} = \sqrt{m!} \bra{l(\btheta)}$ and $\ket{\btheta} = \sqrt{m!}\ket{r(\rev{\btheta})}$ for $\btheta \in \mathbb{R}^m$.}. The $\bfz$ indicates the distributional limit from above; which we will keep implicit from now on. For models without bound states the form factor equations can be summarized as $\{F_{2k}\}_{k\in \mathbb{N}_0}$ being a family of meromorphic functions which satisfy,
\begin{eqnarray}
    &&\text{S-symmetry:}~~ F_{2k}(\bzeta) = S^\tau(\bzeta) F_{2k}(\bzeta^\tau), \; \tau \in \SymGroup_{2k}, \label{eq:ssym}\\
    &&\text{S-periodicity:}~ F_{2k}(\bzeta^\prime,\zeta) = F_{2k}(\zeta,\bzeta^\prime), \label{eq:speriod}
\end{eqnarray}
for $\bzeta = (\bzeta^\prime,\zeta) \in \mathbb{C}^{2k}$ with arbitrary $\bzeta^\prime \in \mathbb{C}^{2k-1}$ and $\zeta \in \mathbb{C}$ and with the following singularity structure: The only poles of all elements of $\{F_{2k}\}$ are first-order and fixed by
 \begin{multline}\label{eq:res}
    \lim_{\zeta'\to \zeta} \, (\zeta-\zeta^\prime) F_{2k+2}(\zeta,\zeta^\prime+i\pi,\bzeta) \\
    = \tfrac{1}{2\pi i} \bigg( 1 - \prod_{p=1}^{k} S(\zeta-\zeta_p) S(\zeta_{p+k}-\zeta) \bigg) F_{2k}(\bzeta)
\end{multline}
as well as consistency with S-symmetry and S-periodicity. These poles are named \emph{kinematical singularities}.

Solutions to these equations take the general form
\begin{eqnarray}\label{eq:ffgenform}
    F^A_{2k}(\bzeta) = q_{2k}(\exp \bzeta) \prod_{1\leq i< j \leq 2k} \frac{ F_\mathrm{min}(\zeta_j-\zeta_i)}{e^{\zeta_i}+e^{\zeta_j}},
\end{eqnarray}
where the $q_{2k}$ are symmetric Laurent polynomials depending on $A$, and $F_{\mathrm{min}}$ is the so-called \emph{minimal solution} of the model which satisfies
\begin{eqnarray}
    F_{\mathrm{min}}(\zeta) &=& S(\zeta) F_{\mathrm{min}}(-\zeta), \\
    F_{\mathrm{min}}(\zeta+i\pi) &=& F_{\mathrm{min}}(-\zeta+i\pi).
\end{eqnarray}
For a conventional normalization, $F_\mathrm{min}(i\pi) = 1$, the minimal solution has an integral representation as
\begin{eqnarray} \label{eq:fint}
    F_\mathrm{min} (\zeta) & = \exp \left(2\int_0^\infty f(t) \sin^2 \frac{(i\pi-\zeta) t}{2\pi} \, \frac{dt}{t\sh t}\right),
\end{eqnarray}
where $f$ is uniquely specified in terms of $S$. For later use we introduce the constant
\begin{eqnarray}
    \finfty = \underset{\theta\to \infty}{\lim} F_{\mathrm{min}}(\theta+i\pi) = \exp \int_0^\infty \frac{f(t)}{t \sh t} dt.
\end{eqnarray}

The form factors of observables in integrable models were subject to a plethora of studies and there exist explicit expressions at low particle numbers in a number of physically relevant models; see e.g. \cite{Smi92,FMS93,CF01,BFK06} . Our aim will be to take these solutions, insert it into \eqref{eq:kernel} to obtain an expression for $\braket{\btheta | A |\bfeta}$ suitable for numerical analysis.

\subsection{Form factors of the stress tensor}\label{ff}

Ultimately, we are interested to treat the smeared stress tensor, i.e., choosing $A = T(\gamma,g^2)$. The stress tensor is mainly characterized as the conserved local hermitian current generating the translations upon spacelike integration,
\begin{equation}\label{eq:stresschar}
    P^\mu = \int T^{\mu 0}(0,x) dx, \qquad \partial^\mu T_{\mu\nu} = 0.
\end{equation}

Poincar\'e covariance of the stress tensor and the second identity in \eqref{eq:stresschar} imply that
\begin{eqnarray}\label{eq:stresstensform}
  F^{T(\gamma,g^2)}_{2k}(\btheta) = u^\mu u^\nu \, t_{\mu\nu}(p(\btheta)) \, F_{2k}(\btheta) \, \tilde{g^2}(u^\rho p_\rho(\btheta)),
\end{eqnarray}
where
\begin{eqnarray}
    \tilde{g^2}(p) & = & \int dx \, g^2(x) e^{ipx},\\
    t_{\mu\nu}(p) & = & \frac{\eta_{\mu\nu} p^2-p_\mu p_\nu}{p^2}, \label{eq:freestresstensor}
\end{eqnarray}
and $\{ F_{2k} \}$ solves the form factor equations (\ref{eq:ssym}--\ref{eq:res}) corresponding to the trace of the stress tensor, $A=\eta^{\mu\nu} T_{\mu\nu}(0)$.

The first identity in \eqref{eq:stresschar} implements a normalization on $F_2$ (see e.g. \cite[Thm.~3.2]{BCM23}),
\begin{eqnarray}
    F_2(i\pi) = \frac{\mass^2}{2\pi},
\end{eqnarray}
fixing also the normalization of higher order $F_{2k}$ via the residue relation \eqref{eq:res}.

\subsection{Explicit solutions for the sinh-Gordon model}\label{shg}

The sinh-Gordon model describes a bosonic scalar massive degree of freedom which scatters according to
\begin{eqnarray}
    S(\zeta) = \frac{\sh \zeta -ia}{\sh \zeta+ia}, \quad 0 < a < 1.
\end{eqnarray}
Here $a=0$ corresponds to the free model and $a=1$ to the maximally interacting case. The scattering function $S$ is obtained in perturbation theory \cite{AFZ79,BS91} from the Lagrangian
\begin{eqnarray}
    \mathcal{L} = \tfrac{1}{2} \partial_\mu \varphi \partial^\mu \varphi - \tfrac{\mass^2}{g^2} \ch g \varphi,
\end{eqnarray}
with $a = \sin \tfrac{\pi B}{2}$ and $B= \frac{2g^2}{8\pi+g^2}$. The minimal solution is given by \eqref{eq:fint} with
\begin{eqnarray}
    f(t) = 4 \sh \tfrac{Bt}{4} \sh \tfrac{(2-B)t}{4} \sh \tfrac{t}{2} (\sh t)^{-1}
\end{eqnarray}
and satisfies
\begin{eqnarray}\label{eq:fminrelshg}
    F_\mathrm{min}(\zeta) F_\mathrm{min}(\zeta+i\pi) = \frac{\sh \zeta}{\sh \zeta + ia} (\finfty)^2.
\end{eqnarray}

The constant $\finfty$ can be evaluated efficiently numerically and interpolates monotonically between $1$ and $\exp(\frac{2 G}{\pi} - \operatorname{acoth}3) \approx 1.26687$ for $a$ ranging from \makebox{0 to 1}. Here $G$ denotes Catalan's constant.

Concerning the form factors of the stress tensor, we first note that $T^{\mu\nu}$ is an even observable with respect to the inversion symmetry $\varphi \leftrightarrow -\varphi$ of the sinh-Gordon model so that odd form factors vanish and the preceding characterization of even form factors applies. Solutions to the form factor equations for $F_{2k}$---representing the trace of the stress tensor---have been obtained before \cite{FMS93,KM93}.
These take the form \eqref{eq:ffgenform} with $q_{2k}$ given by
\begin{eqnarray}
    q_0 &=& 0, \label{eq:q0} \\
    q_2 &=& \frac{m^2}{2\pi} \sigma_1, \label{eq:q2}\\
    q_4 &=& \frac{m^2}{2\pi} \frac{4a}{2\pi} (\finfty)^{-4} \sigma_1 \sigma_2 \sigma_3, \label{eq:q4} \\
    q_6 &=& \frac{m^2}{2\pi} \bigg(\frac{4a}{2\pi}\bigg)^2 (\finfty)^{-12} \sigma_1 \sigma_5\\
    && (\sigma_2\sigma_3\sigma_4 + c_1 (\sigma_4\sigma_5+\sigma_1\sigma_2\sigma_6) + c_2 \sigma_3\sigma_6).
\end{eqnarray}
Here $c_1 = 4(1-a^2)$, $c_2 = 1-c_1$, and $\sigma_j$ denotes the $j$-th elementary symmetric polynomal, i.e., 
$$\sigma_j(x_1,\ldots,x_k) = \quad \sum_{\mathclap{1\leq i_1 < \ldots < i_j \leq k}} \quad  x_{i_1} \ldots x_{i_j}.$$

\subsection{Nonminimal stress tensors}\label{nonmin}

While the form factors given in the previous section correspond to the canonical choice of the stress tensor, other choices are possible. In this section we briefly characterize these options in order to prepare a study of the validity of QEIs for these nonminimal expressions. The motivation here lies in the fact that in the inverse scattering approach one has no direct access to usual prescriptions for obtaining the stress tensor since a Lagrangian or action is a priori not available. While this problem may be overcome in specific models like the sinh-Gordon model, in general, this is difficult and an intrinsic description of the stress tensor in the inverse scattering setup is desirable. In this regard, QEIs may qualify as an additional physical requirement selecting a physically reasonable stress tensor. This has been explored in \cite{BC15,BCM23} at the one-particle level and will be extended here to the two-particle level.

It is convenient to analyze the freedom of choice for the stress tensor at the level of the form factor equations. Based on the decomposition \eqref{eq:stresstensform}, it is sufficient to analyze the freedom of choice of the trace of the stress tensor, which is invariant under boosts. This implies that $F_{2k}$ is invariant under uniform shifts, i.e., $F_{2k}(\btheta+(\lambda,\ldots,\lambda)) = F_{2k}(\btheta)\text{ for all }\btheta \text{ and }\lambda,$ constraining $q_{2k}$ from \eqref{eq:ffgenform} to be homogeneous of degree $k(2k-1)$. Restricting to polynomial $q_{2k}$ and following the analysis in \cite{KM93}, a general solution up to the two-particle level is of the form \eqref{eq:ffgenform} with
\begin{eqnarray}
    q_4 = \frac{m^2}{2\pi} \frac{4a}{2\pi} \left( (1-\lambda) \sigma_1 \sigma_2 \sigma_3 + \lambda (\sigma_4 \sigma_1^2 + \sigma_3^2)\right), \quad \lambda \in \mathbb{R}
\end{eqnarray}
and $q_2$ as given in \eqref{eq:q2}; higher $q_{2k}, k \geq 3$ do not contribute at the two-particle level. As outlined in \cite[Sec.~4.3]{FMS93} the only choice which yields a singularity structure compatible with the form factor equations for any component of the stress tensor $u^\mu u^\nu T_{\mu\nu}$ corresponds to $\lambda = 0$; the canonical choice for the stress tensor as given in the preceding section.

Other freedoms include to consider nonzero real constants $q_0$ or $q_1$; the latter case has been analyzed in \cite{MS94}. Both result in constant shifts of the spectrum of $T(\gamma,g^2)$ with a possible dependence on the type of states: For $q_0$, we simply have $c_g \to c_g + q_0 \int d\tau g^2(\tau)$ in \eqref{eq:introQEI}. For states with fixed particle number $q_1$ does not contribute to $c_g$. Thus both alterations can be accomodated for in a simple manner and we assume $q_0=q_1=0$ in the following.

The only remaining freedom of choice is to ``add derivatives'': Given the canonical solution represented by $\{ q_{2k} \}$, multiplication by an arbitrary power of $\sigma_{2k-1} \sigma_1 / \sigma_{2k}$ at each $k\in\mathbb{N}$ alters neither the form factor equations nor \eqref{eq:stresschar} and has the same degree of homogeneity as $q_{2k}$. Thus, an arbitrary solution will be of the form $\{ r(\sigma_{2k-1} \sigma_1 / \sigma_{2k}) q_{2k} \}$ where $r$ is an arbitrary polynomial with real coefficients and $r(0)=1$. At the operator level, this corresponds to modifying the canonical stress tensor $T^{\mu\nu}$ into $r(-\square /m^2) T^{\mu\nu}$, where $\square = \partial_\mu \partial^\mu$ denotes the d'Alembert operator. For the special case that the degree of $r$ is $1$, we may parametrize our nonminimal stress tensor to be of the form
\begin{eqnarray}\label{eq:nonmin1}
    T^{\mu\nu}_{\mathrm{nonmin},\nu} = \left( (1-\nu) - \nu \square \right) T^{\mu\nu},
\end{eqnarray}
where $T^{\mu\nu}$ denotes the canonical expression for the stress tensor as before.

\section{Numerical setup}\label{setup}

In this section we will motivate and explain the numerical setup. Our aim is to approximate an optimal bound $c_g$ as appearing in \eqref{eq:introQEI}. The first step is to discretize the hermitian quadratic form $A_k=\proj_{S}^{(k)} T(\gamma,g^2) \proj_{S}^{(k)}$ acting on $L^2_S(\mathbb{R}^k)$. Next, we will perform a numerical spectral analysis of the resulting matrix in order to find its lowest eigenvalue. If the resolution of the discretized problem is sufficiently high, we expect to obtain a good approximation for $c_g$. Note that it is sufficient to study $A_k$ on $L^2_<(\mathbb{R}^k):=L^2(\mathbb{R}^k, d\btheta_{\theta_1 < \ldots < \theta_k})$, since $L^2_S(\mathbb{R}^k) = \proj_S^{(k)} L^2_<(\mathbb{R}^k)$.

\subsection{Physical parameters}\label{physical}
There are four physically relevant inputs to the analysis: The class of states, the interaction model, the observer's trajectory and the averaging profile. The class of states consists of all (normalized) states with particle number $k$; in this case being either one or two. The interaction model is the sinh-Gordon model specified by its coupling parameter $B \in [0,2]$ and its mass scale $m > 0$. For the mass, we choose to measure energy in rescaled units so that effectively $m=1$. Note here that the mass scale of the stress tensor is fixed by \eqref{eq:stresschar} to $m^2$ and that the form factors of the (trace of the) stress tensor $F_{2k}$ are all proportional to $\mass^3$ since we have factored out the Fourier transformed averaging function $\tilde{g^2}$ which has mass scale $m^{-1}$.

For inertial trajectories $\gamma$ we may assume without loss of generality, that $\gamma(\tau)= (\tau,0)$: By covariance of the stress tensor, expectation values transform as
\begin{equation}
    \braket{\psi, T(\gamma,g^2) \psi} \to \braket{\psi', T(\gamma',g^2)\psi'}, 
\end{equation}
where the primes indicate the transformed objects. For a generic inertial trajectory $\gamma$ there is a transformation including a boost and a translation which brings it into the form $\gamma': \tau \mapsto (\tau,0)$. The boost results in a shift of $\psi$; the translation multiplies it with a phase factor. Since we optimize over arbitrary states (aside from fixing the particle number), we can ignore the change from $\psi$ to $\psi'$ without loss of generality. The introduced cutoff for the numerical analysis does not affect this argument since we choose it sufficiently large for ignoring boundary effects.

Concerning the profile, we will study two one-parameter classes of smooth test functions which satisfy $\int g^2(x) dx = 1$: A Gaussian profile,
\begin{eqnarray}
    g_\mathrm{G}(x) = (4\pi\scale^2 )^{- \tfrac{1}{4}} \exp(-\tfrac{x^2}{8\scale^2}), \; & \tilde{g_\mathrm{G}^2}(p) = \exp(-\scale^2 p^2), \nonumber
\end{eqnarray}
and a Lorentzian one,
\begin{eqnarray}
    g_\mathrm{L}(x) = (2\scale)^{\tfrac{1}{2}} \exp(-\tfrac{|x|}{2\scale}),\; & \tilde{g_\mathrm{L}^2}(p) = \frac{1}{1+\scale^2 p^2}. \nonumber
\end{eqnarray}
Both choices are parametrized by a single parameter, the averaging scale $\scale$ which indicates the time scales over which $A_k$ is averaged.

\subsection{Discretization method}\label{discretization}
For resolution $n\in\mathbb{N}$ and rapidity cutoff $R> 0$, we define a lattice spacing $h := \frac{2R}{n}$, lattice points $\lambda_j := -R + h \cdot (j-\tfrac{1}{2})$, and an ordered index set 
$$I^k_{n,<} := \{ \bm{j} \in \{ 1,\ldots,n\}^k : \; j_1 < \ldots < j_k \},$$
enumerating a hyper triangular lattice
\begin{eqnarray*}
    \Lambda_{n,R,<}^k & = & \{ \blambda_{\bm{j}}:=(\lambda_{j_1}, \ldots, \lambda_{j_k}) \}_{\bm{j} \in I^k_{n,<}}.
\end{eqnarray*}
Then, towards large $n$ and $R$, the orthonormal system
\begin{eqnarray}
    E^{n,R} & = &\{ e^{n,R}_{\bm{j}} \}_{\bm{j} \in I^k_{n,<}}, \quad e^{n,R}_{\bm{j}}(\btheta) := \chi_{j_1}(\theta_1) \ldots \chi_{j_k}(\theta_k), \nonumber
\end{eqnarray}
given in terms of normalized characteristic functions
\begin{eqnarray}
    \chi_j(\theta) & = & h^{-\tfrac{1}{2}} \left\lbrace \begin{matrix} 1, & \quad \theta \in  \lambda_j+[-h/2,h/2],\\ 0, & \quad \text{otherwise} \end{matrix} \right. \nonumber
\end{eqnarray}
decently approximates $L^2_<(\mathbb{R}^k)$ and thus $L^2_S(\mathbb{R}^k)$. In particular, for $\psi \in L^2_S(\mathbb{R}^k)$ we have
\begin{eqnarray}
    \braket{\psi, A \psi} &=& (k!)^2 \sum_{\bm{i},\bm{j} \in I^k_{n,<}} \beta_{\bm{i}}^\ast M^A_{\bm{i},\bm{j}} \beta_{\bm{j}},
\end{eqnarray}
where
\begin{eqnarray}
    \beta_{\bm{j}} & =& \braket{e_{\bm{j}},\psi} \text{ and } M^A_{\bm{i},\bm{j}} =  \braket{e_{\bm{i}}, A \, e_{\bm{j}}}.
\end{eqnarray}
The matrix elements evaluate to
\begin{eqnarray}\label{eq:matrixelement}
     M^A_{\bm{i},\bm{j}} &=& h^k \int_{\big[-\tfrac{1}{2},\tfrac{1}{2}\big]^{2k}} d\btheta d\bfeta \, A(\blambda_{\bm{i}} + h \btheta , \blambda_{\bm{j}}+h\bfeta ).
\end{eqnarray}
Here, hermiticity of $A$, thus of $M^A$, corresponds to
\begin{eqnarray}
    A(\btheta,\bfeta) = \overline{A(\bfeta,\btheta)} \quad \leftrightarrow \quad M^{A}_{\bm{i},\bm{j}} = (M^A_{\bm{j},\bm{i}})^*.
\end{eqnarray}

In order to efficiently compute the components of $M^A$, we will approximate it for small $h$ by its expansion up to $\mathcal{O}(h^k)$, resp., by expanding the integrand $A(\blambda_{\bm{i}}+h\btheta,\blambda_{\bm{j}}+h\bfeta)$ up to $\mathcal{O}(h^0)$. If $i_r \neq j_s$ for all $r,s \in \{1,\ldots,k\}$, then $A(\blambda_{\bm{i}},\blambda_{\bm{j}})$ is regular and we may simply use a leading order approximation for small $h$,
$$(A(\blambda_{\bm{i}} + h \btheta , \blambda_{\bm{j}}+h\bfeta ))_{\mathrm{l.o.}} = A(\blambda_{\bm{i}},\blambda_{\bm{j}}),$$
where the subscript ``l.o.'' indicates ``leading order''. The expression is independent from $\btheta$ and $\bfeta$ so that
\begin{eqnarray}
    (M^A_{\bm{i},\bm{j}})_{\mathrm{l.o.}} = h^2 A(\lambda_{\bm{i}},\lambda_{\bm{j}}), \quad i_r \neq j_s \text{, } _{r,s\in \{1,\ldots ,k\}}.
\end{eqnarray}

However, whenever $i_r=j_s$ for one or more combinations of $(r,s)$, we have to deal with delta-distributions and kinematical singularities; cf. \eqref{eq:kernel} and \eqref{eq:res}. This results in a directional dependence, i.e., a dependence on $\btheta,\bfeta$, which has to be integrated over according to \eqref{eq:matrixelement} and the necessity to compute next-to-leading order contributions in some cases.

\subsection{Computation of leading order contributions to the discretized kernels}\label{explcomp}
To illustrate the computation of the leading-order contribution to $M^A$ for the singular case, we treat here two exemplaric cases in detail. Results for the other cases are given in the next two sections. Since in the one-particle case $A(\lambda_i,\lambda_j)$ is regular (even for coinciding arguments $\lambda_i=\lambda_j$), we focus on the two-particle case, i.e., $k=2$. The subscript ``l.o.'' indicates again the leading-order contribution for small $h$.

Let us introduce the short hand notation $S_{ij} := S(\lambda_i - \lambda_j)$ and $F_{k,\bm{i}\bm{j}} := F_k(\blambda_{\bm{i}},\blambda_{\bm{j}}+i\bpi)$. Let us also adopt the convention that $\ch_{i} = \ch \lambda_i$, $\ch_{ij} = \ch(\lambda_i-\lambda_j)$ and similarly for $\sh_i$ and $\sh_{ij}$. 

Suppose that $i_1 \neq j_1$ and $i_2=j_2$, then either by direct computation from the expression for $F_4$ (\cref{shg}) or according to \eqref{eq:res}, we obtain that
\begin{multline}
    (F_4(\blambda_{\bm{i}}+h\btheta,\blambda_{\rev{\bm{j}}}+h\rev{\bfeta}+i\bpi))_{\mathrm{l.o.}} \\
    = h^{-1} \frac{\big( 1 - S_{i_2i_1} S_{j_1j_2} \big) F_{2,i_1j_1}}{2\pi i (\theta_2-\eta_2+i0)}.
\end{multline}
Since 
\begin{eqnarray}
    \int_{\big[-\tfrac{1}{2},\tfrac{1}{2}\big]^{4}} \frac{d\btheta d\bfeta}{2\pi i (\theta_2-\eta_2+i0)} = -\frac{1}{2},
\end{eqnarray}
we arrive at
\begin{eqnarray}
    (M_{\bm{i},\bm{j}}^{F_4})_{\mathrm{l.o.}} = \frac{h}{2} \bigg(S_{i_2 i_1} S_{j_1j_2} - 1 \bigg) F_{2,i_1j_1}, \quad \substack{i_1 \neq j_1\\i_2=j_2}.
\end{eqnarray}

Suppose instead that $\bm{i} = \bm{j}$, then
\begin{multline}
    (F_4(\blambda_{\bm{i}}+h\btheta,\blambda_{\rev{\bm{j}}}+h\rev{\bfeta}+i\bpi))_{\mathrm{l.o.}} \\
    = \frac{2a\ch_{i_1 i_2}}{a^2 + \sh^2_{i_1i_2}} \frac{((\theta_1-\eta_1)\ch_{i_1} + (\theta_2-\eta_2) \ch_{i_2})^2}{4\pi^2(\theta_1-\eta_1+i0)(\theta_2-\eta_2+i0)};
\end{multline}
again by direct computation. The terms of the form \makebox{$(\theta_1-\eta_1)(\theta_2-\eta_2+i0)^{-1}$} vanish upon integration, so that only the term which does not depend on $\btheta$ and $\bfeta$ survives:
\begin{eqnarray}
    (M_{\bm{i},\bm{j}}^{F_4})_{\mathrm{l.o.}} = h^2 \frac{a}{\pi^2} \frac{\ch_{i_1} \ch_{i_2}\ch_{i_1 i_2}}{a^2 + \sh^2_{i_1i_2}}, \quad \bm{i}=\bm{j}.\quad
\end{eqnarray}

Expressions for $F_2(\lambda_{i_1}+h\theta_1,\lambda_{j_1}+h\eta_1+i\pi) \delta(\lambda_{i_2}-\lambda_{j_2} +h(\theta_2-\eta_2))$ and permuted variants can be computed in analogous fashion. Here one uses that $\delta(\lambda+h\theta) = h^{-1} \delta(\theta)$ if $\lambda = 0$ and $\delta(\lambda +h\theta) = 0$ in case that $\lambda \neq 0$ for sufficiently small $h$.

\subsection{Explicit discretized kernels up to leading order}\label{explexpr}
In this section, we will list the discretized expressions obtained by the method described in the preceding section focussing on the leading order terms. From \eqref{eq:kernel} we obtain schematically that
\begin{eqnarray}
    A_1 &=& F_2, \\
    A_2 &=& \tfrac{1}{2} ( F_4 + 4 F_2 \delta_2) ,
\end{eqnarray}
where the projectors $\proj^{(k)}_{S,\btheta}$ and $\proj^{(k)}_{S^{-1},\bfeta}$ are included in the definition of the summands. We will list the matrix elements obtained for the individual summands \emph{up to leading order} in $h$:
\begin{eqnarray}
    (M_{i,j}^{F_2})_{\mathrm{l.o.}} & = & h F_{2,ij},  \label{eq:f2lo}
\end{eqnarray}
\begin{eqnarray}
    (M_{\bm{i},\bm{j}}^{F_4})_{\mathrm{l.o.}} & = & \left\lbrace \begin{matrix}
    h^2 \frac{a}{\pi^2} \frac{\ch_{i_1} \ch_{i_2} \ch_{i_1 i_2}} {a^2+\sh^2_{i_1i_2}} & \quad \bm{i} =\bm{j} \\
    h^2 \frac{a}{\pi^2} S_{i_2i_1} \frac{\ch_{i_1} \ch_{i_2} \ch_{i_1 i_2}}{a^2+\sh^2_{i_1i_2}} & \quad \bm{i} =\rev{\bm{j}}\\
    h\tfrac{1}{2} (1-S_{i_2 i_1}S_{j_1j_2}) F_{2,i_2j_2} & \quad i_1 =j_1 \\
    h\tfrac{1}{2} (S_{j_1 j_2}-S_{i_2 i_1}) F_{2,i_2j_1} & \quad i_1 =j_2 \\
    h\tfrac{1}{2} (S_{i_2i_1}-S_{j_1j_2}) F_{2,i_1j_2} & \quad i_2 =j_1\\
    h\tfrac{1}{2} (S_{i_2i_1} S_{j_1 j_2}-1) F_{2,i_1j_1} & \quad i_2 =j_2 \\
    h^2 F_{4,\bm{i}\rev{\bm{j}}}& \quad\text{otherwise}  \end{matrix} \right. , \label{eq:f4lo}
\end{eqnarray}
\begin{eqnarray}
    (M_{\bm{i},\bm{j}}^{4 F_2 \delta_2})_{\mathrm{l.o.}} &=& \left\lbrace \begin{matrix}
    h\frac{1}{2\pi} (\ch^2_{i_1} + \ch^2_{i_2})  & \quad \bm{i}=\bm{j} \\
    h\frac{1}{2\pi} S_{i_2i_1} (\ch^2_{i_1} + \ch^2_{i_2}) & \quad \bm{i} = \rev{\bm{j}} \\
    h S_{i_2i_1} S_{j_1 j_2} F_{2,i_2j_2}& \quad i_1=j_1 \\
    h S_{i_2i_1} F_{2,i_2j_1} & \quad i_1=j_2 \\
    h S_{j_1 j_2} F_{2,i_1j_2}& \quad i_2=j_1\\
     h F_{2,i_1j_1}& \quad i_2=j_2 \\
    0 & \quad \text{otherwise} \end{matrix} \right. . \label{eq:f2delta2lo}
\end{eqnarray}
Here, $F_{2,ij} = \frac{1+\ch(\lambda_i+\lambda_j)}{4\pi} F_\mathrm{min}(\lambda_{i}-\lambda_{j}+i\pi)$. The case distinctions are concatenated by ``else if'', so that reading from top to bottom the first affirmative condition is accepted. For instance, the third case in $M^{F_4}_{\bm{i},\bm{j}}$ corresponds to $i_1=j_1 \wedge i_2 \neq j_2$ since if $i_2=j_2$ the first condition would have already been true. Note that, when a single pair of arguments is equal, the contributions from $F_4$ and $F_2\delta_2$ are of a similar form, so that they add up nicely. Note further that (\ref{eq:f2lo}--\ref{eq:f2delta2lo}) are also valid for nonminimal stress tensors as given in \eqref{eq:nonmin1} when appropriate factors are added to $F_{2,ij}$ and $F_{4,\bm{i}\rev{\bm{j}}}$.

\subsection{Next-to-leading order contributions}\label{explexprnlo}

In the numerical analysis, we aimed to include all terms up to order $\mathcal{O}(h^2)$. Since in the singular cases the leading-order contribution of $M^A$ is $\mathcal{O}(h)$ this implies that $M^A$ receives next-to-leading order corrections in the cases $i_1=j_1$, $i_1=j_2$, $i_2=j_1$ and $i_2=j_2$, for both $A=F_2 \delta_2$ and $A=F_4$. In the special cases $\bm{i}=\bm{j}$ and $\bm{i}=\rev{\bm{j}}$ only $A=F_2 \delta_2$ receives next-to-leading order contributions. At least some of these terms are significant: Without inclusion, previous tests showed a bad numerical convergence in small $h$ (i.e. with increasing $n$) which was drastically improved when including the terms mentioned above. The discretized kernels including next-to-leading order contributions are too long to be displayed here but arise from tedious but straightforward expansion of previously given expressions in higher orders of $h$. For faster computations we neglect terms which depend on derivatives in $F_\mathrm{min}$. Their contribution has been checked to be small for a large range of parameters as is also indicated by the smallness of $\lvert F_\mathrm{min}^{\prime}/F_\mathrm{min} \rvert$.

\subsection{Numerical representation of the minimal solution}\label{numminsol}
It is desirable to have a fast numerical representation of the minimal solution $F_\mathrm{min}$ which appears frequently in the expressions computed above. The minimal solution is defined via an integral transform according to \eqref{eq:fint}. We implemented this integral transform numerically and used the methods suggested in \cite[Eq.~\lParen 4.18\rParen , $N=1$]{FMS93} to improve the convergence rate. The relation \eqref{eq:fminrelshg} is satisfied with a precision of $10^{-6}$. We stored the numerical approximation in a lookup table for fast evaluation.

\subsection{Spectral Analysis}\label{spectral}
The eventual task is to find the lowest eigenvalue of the dense Hermitian matrix $M^A$. The corresponding vector space dimension is given by the number of lattice points and for large $n$ amounts to
$$d = | I_{n,<}^k| = \left( \begin{matrix} n \\ k \end{matrix}\right) \approx \frac{n^k}{k!}.$$

We use Lanczos', resp., Arnoldi's method as implemented in Mathematica (Version 13.2) in the function \includegraphics{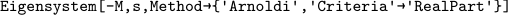} The outcome of this analysis will of course depend not only on the physical parameters (\cref{physical}) but also on $n$ and $R$. Thus the result is only accepted as an approximate optimal bound $c_g$ whenever it stabilizes for sufficiently large $n$ and $R$ upon further variation of $n$ and $R$ within a satisfactory precision goal.

\subsection{Plausibility checks and error bounds}\label{plausibility}
There are a number of plausibility checks available. We have checked that the resulting matrices are hermitian (they have real eigenvalues) and that the discretization and spectral analysis method reproduces the spectrum of known singular integral operators like the Hilbert transform. We have also checked that for the free scalar model (i.e., $a=0$ in the sinh-Gordon model) we obtain positivity in one- and two-particle states ($k=1,2$). Moreover, the one-particle results (i.e., looking at $A_1$) agree with those obtained in \cite{BC15}. Importantly for two-particle results, the numerical factor in front of $F_4$ is fixed by the residue relation \eqref{eq:res} and can be confirmed by direct computation.

An indication for the size of numerical errors (disregarding systematical ones) arises from the precision goal mentioned in \Cref{spectral}: All finite spectral bounds have been checked to stabilize with respect to $n$ and $R$ (where $R$ is chosen close to the support of the lowest eigenvector). Upon variation of $R$ within 10 $\%$ margins, the spectral bounds varied within two relative digits around $n = 50$ and within five relative digits around $n=100$; as is also indicated by the fast decay of the eigenstates towards the boundary (Sec.~\ref{states}). For variation of $n$ within a 10 $\%$ margin we obtain the same precision at the one-particle level. For the two-particle case we indicate convergence in $n$ explicitly in the following plots.

\section{Numerical results on Negative Energies}\label{results}
In this section we summarize our numerical results, presenting approximately optimal lower bounds on the spectrum of the smeared stress tensor $T(\gamma,g^2)$ within the class of one- and two-particle states for the sinh-Gordon model. The results at the one-particle level are in precise agreement with a previous analysis in \cite{BC15}. The results for the two-particle level are new.

Throughout this section, we will specify the inputs of the numerical analysis. This includes physical parameters (\cref{physical}): the coupling constant $B \in [0,2]$ fixing the interaction and the averaging function (default: Gaussian, alternative: Lorentzian) with averaging scale $\scale$ within $10^{-3} \,m^{-1} \ldots 1.0 \,m^{-1}$. It further includes numerical parameters (\cref{discretization}): The number of lattice points per side $n$ and the rapidity cutoff $R$. We restrict our discussion to inertial trajectories which without loss of generality may be considered to take the form $\gamma(\tau) = (\tau,0)$ (\cref{physical}). In this case, we write
$$ T^{00}(g^2) \equiv \int d\tau \, T^{00}(\tau,0) g^2(\tau) = T(\gamma,g^2).$$
Since we restrict our spectral analysis to the class of one- and two-particle states, we effectively minimize $T^{00}_k (g^2) \equiv \proj^{(k)} T^{00}(g^2) \proj^{(k)}$ with $k=1,2$.

As outcomes of our numerical evaluation, we find eigenstates minimizing energy expectation values (\cref{states}) and corresponding estimates for an optimal lower bound $c_g$ of QEI form \eqref{eq:introQEI}; including its dependence on interaction strength  (\cref{coupling}) and on the averaging scale (\cref{averaging}). We also analyze constraints on nonminimal expressions for the stress tensor imposed by the validity of QEIs (\cref{nonminimal}).

\subsection{Lowest eigenstates}\label{states}
\begin{figure}[h!]
    \begin{minipage}{0.235\textwidth}
    \includegraphics[width=\linewidth]{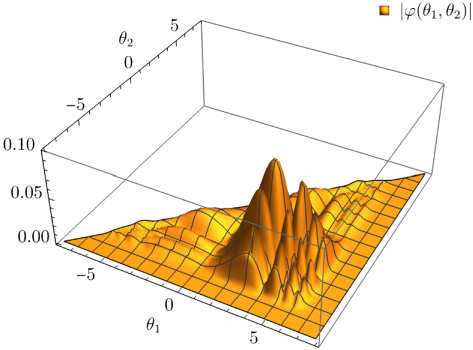}
    \end{minipage}
    \hspace{\fill} 
    \begin{minipage}{0.235\textwidth}
    \includegraphics[width=\linewidth]{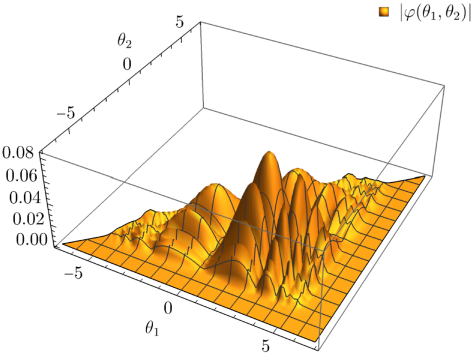}
    \end{minipage}

    \vspace*{.1cm} 

    \begin{minipage}{0.235\textwidth}
    \includegraphics[width=\linewidth]{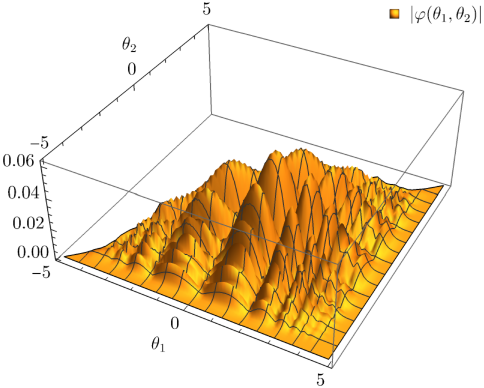}
    \end{minipage}
    \hspace{\fill} 
    \begin{minipage}{0.235\textwidth}
    \includegraphics[width=\linewidth]{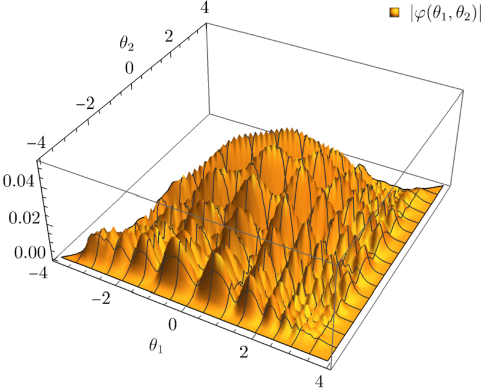}
    \end{minipage}

    \caption{Absolute value of lowest eigenvector $|\varphi(\theta_1,\theta_2)|$ of $T^{00}_2(g^2)$ for the maximally interacting sinh-Gordon model ($B=1$). Parameters: $n=80$ and, from upper left to lower right, $(R,\scale)=(8.0,0.008 \,m^{-1})$, $(7.0,0.021 \,m^{-1})$, $(5.0,0.086 \,m^{-1})$, $(4.0,0.220 \,m^{-1})$.}
    \label{fig:lestateshg}
\end{figure}

In this section we plot the most negative eigenstate $\varphi$ of $T_2^{00}(g^2)$ for the maximally interacting sinh-Gordon model and varying averaging scale (Fig.~\ref{fig:lestateshg}). In rapidity space, we plot $(\theta_1,\theta_2) \mapsto |\varphi(\theta_1,\theta_2)|$ restricted to the region $\theta_1 \geq \theta_2$. This captures the essential features of $\varphi$ as $|\varphi|$ is symmetric, i.e., $|\varphi(\theta_1,\theta_2)| = |\varphi(\theta_2,\theta_1)|$ due to S-symmetry and unitarity of $S$.

To begin with, we observe that all eigenstates in Figure~\ref{fig:lestateshg} are narrowly localized showing fast decay towards the boundary indicating convergence of the numerical analysis. In contrast, for unbounded/nonconvergent cases computed eigenstates have been found to be localized predominantly at the boundary indicating divergence with increasing $R$. In comparison to the lowest one-particle eigenstates $\varphi_1$ \cite{BC15}, we find that the values of $\varphi$ are bigger in magnitude leading to more negative bounds (\cref{coupling}).

Restricting to large averaging scales or regions with sufficiently far separation between $\theta_1$ and $\theta_2$ (front quarter), profiles at constant $\theta_1$ (or $\theta_2$) share some similarities with the one-particle profiles $|\varphi_1|$. This goes in line with the expectation that in these cases states should behave like a tensor-product of two one-particle states---as in the free theory---since $S(\theta) \to 1$ for $\theta \to 0$, resp., $\theta \to \infty$.

\subsection{Dependence on the interaction strength}\label{coupling}

\begin{figure}[h!]
    \includegraphics{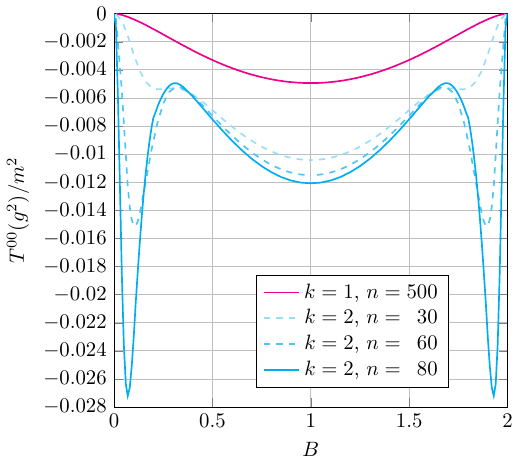}
    \caption{Lowest eigenvalue of $T_k^{00}(g^2)$ for varying coupling constant $B$ in the sinh-Gordon model. We compare the lowest eigenvalues in one- and two-particle states and indicate its variance for increasing $n$ for the latter. Parameters: $R=4.5$ or $5.0$, $\scale=0.1 \,m^{-1}$.}
    \label{fig:levcouplingshg}
\end{figure}

In this section we analyze the dependence of the lowest eigenvalue of $T^{00}(g^2)$ on the interaction strength as represented by the coupling constant $B$ ranging from $0$ (non-interacting) via $1$ (maximally interacting) through to $2$ (non-interacting) and taking into account the symmetry of the model, $B \leftrightarrow 2-B$; confer Figure~\ref{fig:levcouplingshg}.

Despite the larger numerical uncertainties in the boundary of the graph $0.0 < B < 0.3$ ($1.7 < B < 2.0$), we see that for the free limit $B \to 0$ ($B \to 2$) all curves tend towards zero. This is the expected non-negativity of the energy density in the massive free scalar model for states with fixed particle number. Further, in the region $0.3 < B < 1.7$ we find a qualitatively similar behaviour of the curves in the one- and two-particle case; the eigenvalues decrease monotonously, reaching a local minimum at the maximally interacting case $B=1$. This aligns well with the general expectation that self-interaction is causing the presence of negative energies and that strong self-interaction allows for a stronger accumulation of negative energy. Notable is that the minimum for the two-particle case is more than twice as negative as the one-particle miminum with minima evaluating to $c_g \approx 12 \times 10^{-3}\,m^2$, resp., $c_g \approx 5 \times 10^{-3}\,m^2$; indicating that increasing the particle number allows for a stronger accumulation of negative energy.

Surprisingly, at the two-particle level in the region $0 < B < 0.3$ ($1.7 < B < 2.0$) we find a feature which was not at all visible in the one-particle case. Here the energy density can become much more negative than in the one-particle case and also compared with the stronger interaction regime $0.3 < B < 1.7$; a mininum for the energy density is obtained at about $B \approx 0.07$ with $c_g \approx 2.7 \times 10^{-2} \,m^2$. This finding is very surprising since on general grounds a strong interaction should allow for a more negative energy density and the author is not aware of any discontinuity in the coupling constant for the sinh-Gordon model (within the mentioned regime). While we emphasize that we have run a number of plausibility checks as illustrated in \Cref{plausibility}, at present we cannot exclude this feature to be an artifact of the numerics as is also indicated by the huge variance with respect to $n$. A further increase of $n$ or an expansion of the discretized stress tensor in higher orders of $h$ is desirable and expected to lead to higher precision but is severely limited due to a rapid increase in memory and runtime complexity (\cref{spectral}).

\subsection{Dependence on the averaging scale}\label{averaging}

\begin{figure}[h!]
    \includegraphics{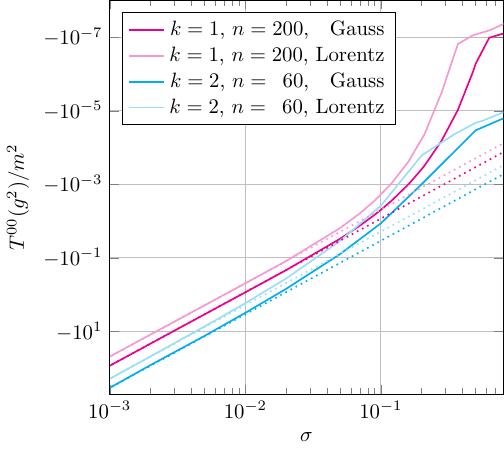}
    \caption{Lowest eigenvalue of $T^{00}_k(g^2)$ for varying averaging scale $\scale$ in the maximally interacting sinh-Gordon model ($B=1$) at the one- and two-particle level for Gaussian and Lorentzian averaging. Parameters: $n=60$, and adaptively chosen $R$. Dotted lines indicate the asymptotic scaling ($\propto \scale^{-2}$).}
    \label{fig:shgavg}
\end{figure}

In this section we plot the dependence of the lowest eigenvalue of $T^{00}(g^2)$ on the averaging profile and scale (Fig.~\ref{fig:shgavg}). For the averaging profile we consider a Gaussian and a Lorentzian smearing function both parameterized by an averaging scale parameter $\scale$ (\cref{physical}), indicating the time scales over which the energy density is averaged.

We find that, the choice of averaging profile basically yields a horizontal shift of the plot linked to the slower decay properties of the Lorentzian profile and that there is a strong dependence on the averaging scale showing universal features. For small $\scale$, i.e., for narrow time spans, we confirm an asymptotic scaling like $\scale^{-2}$ (dotted lines). This is expected since $T^{00}(g^2)$ has scaling dimension 2 and was also confirmed for one-particle states in \cite{BC15}. For large $\scale$, i.e., for long averaging times, the eigenvalues tend to zero rapidly. Note here that the numerics becomes unstable for eigenvalues with absolute values below $\approx 10^{-7} \, m^{-1}$ or for $\sigma$ being larger than $\approx 0.5 \, m^{-1}$.

Despite this, we see clear evidence that, within our numerical precision, the (timelike) averaged energy density becomes non-negative in the long-time limit $g \to 1$; implying the validity of the averaged weak energy condition (AWEC) \eqref{eq:awec} in one- and two-particle states. We also see that in the two-particle case and for Gaussian profile it requires larger averaging scales to approach non-negativity than in the one-particle case and for Lorentzian profile.

\subsection{Nonminimal stress tensors}\label{nonminimal}

In this section we analyze the validity of QEIs for nonminimal expressions of the stress tensor. These arise when adding terms of the form $c \, \square^n T^{\mu\nu}$ to $T^{\mu\nu}$ for some $c \in \mathbb{R}$ and $n\in\mathbb{N}$.

\begin{figure}
    \includegraphics{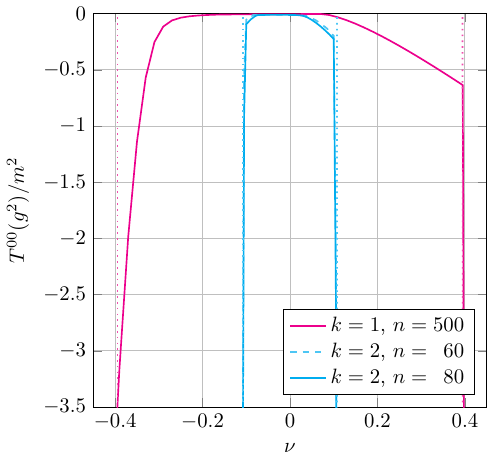}
    \caption{Lowest eigenvalue of $T^{00}_{k,\mathrm{nonmin},\nu}(g^2)$ for nonminimal choice of the stress tensor as in \eqref{eq:nonmin1} in the maximally interacting sinh-Gordon model ($B=1$). We compare the lowest eigenvalues in one- and two-particle states for variying $\nu$ and indicate convergence with increasing $n$ for the latter. The asymptotes of the curves are indicated by dotted lines. Parameters: $R=4.5$ or $5.0$, $\scale=0.1 \,m^{-1}$.}
    \label{fig:nonminimal1}
\end{figure}

When $n > 1$, we have checked this to lead to divergent bounds for arbitrarily small coefficients. This is in agreement with the results at one-particle level obtained in \cite{BC15}. For $n=1$, we parameterize our family of nonminimal expressions for the stress tensor as in \eqref{eq:nonmin1}; denoted by $T^{\mu\nu}_{\mathrm{nonmin},\nu}$ with parameter $\nu$. A plot of the lowest eigenvalues of $T_{k,\mathrm{nonmin},\nu}^{00}(g^2)$ is given in Figure~\ref{fig:nonminimal1}.

For both, one- and two-particle states, we find a $\nu$-band around the canonical choice $\nu = 0$ with finite eigenvalues which are relatively constant with respect to $\nu$. It is notable that the eigenvalues are not symmetric around $\nu = 0$. At the boundary of the bands, the eigenvalues start to decrease rapidly and the eigenstates change appearance. In contrast to the full plot, the position of these thresholds is symmetric around $\nu=0$ and given as dotted vertical lines in the plot. For the one-particle case the threshold lies at $|\nu| \approx 0.395 \approx \frac{1}{2 \finfty}$ in agreement with \cite{BC15} (the r.h.s. corresponds to the analytic result given in the reference). For the two-particle case the threshold for $|\nu|$ is between $0.10$ and $0.11$ (in this case no analytic result is known).

To sum it up, we find that at higher particle numbers the validity of a QEI becomes more restrictive. Speculating on this observation and taking into account the significantly lower bounds in the two-particle case, this indicates that the threshold decreases even more at higher particle numbers; possibly selecting the canonical stress tensor uniquely in states of arbitrary particle number.

\section{Conclusion}\label{conclusion}

In summary, this article introduces a novel numerical method for determining optimal QEI bounds in one- and two-particle states, extending previous work focused on the one-particle case. This extension encountered two formidable challenges: a substantial increase in complexity, affecting runtime, memory, and analytical considerations, and the necessity for numerical treatment of kinematical singularities not encountered in the one-particle domain. The proposed method is inherently applicable to all integrable models constructed via the inverse scattering approach, provided that solutions to the form factor equations are well understood.

We applied our method to a representative example, the sinh-Gordon model, gaining explicit insights into its optimal QEI bounds. A key revelation is the preservation of QEIs at the two-particle level, accompanied by a notable increase in the negativity of energy density compared to the one-particle case. Additionally, our findings substantiate that self-interaction acts as the source of negative energy, with stronger interactions yielding more pronounced negativities. Despite this, we found strong evidence that negative energies can not be sustained for arbitrarily long times which implies the validity of the averaged weak energy condition (AWEC) at both one- and two-particle levels. Last but not least, we established the existence of a one-parameter class of nonminimal stress tensors satisfying QEIs at both levels, with more stringent constraints arising from two-particle QEIs.

The transition from the one-particle to the two-particle level was marked by significant advancements. States featuring a single particle may overlook certain aspects of self-interaction, and the inclusion of two-particle states introduces entirely new technical challenges related to the kinematical singularities. While complexity escalates when considering particle numbers higher than two, no fundamentally new aspects emerge, suggesting a possible continuity in qualitative features. From that perspective our observations stemming from one- and two-particle analyses suggest that QEIs and the AWEC may extend to arbitrary finite particle numbers. Furthermore, we expect that constraints on nonminimal expressions for the stress tensor imposed by the validity of a QEI become more restrictive with increasing particle numbers promising to establish QEIs as a robust selection principle for physically reasonable stress tensors. To confirm these speculations, it is desirable addressing states with higher particle numbers, albeit necessitating novel numerical approximations or additional analytic insights, given the escalating complexity for particle numbers beyond two.

Another important aspect yet to be explored is the treatment of QEIs in states beyond fixed particle numbers. After all, it is well known that---even in non-interacting models---negative energies arise when superposing states with different particle numbers. Thus, treating negative energy states, exemplified by superpositions of vacuum, one- and two-particle states, in a setup with self-interaction presents an intriguing challenge which is within the scope of the current method. We anticipate returning to this issue in future investigations.

\begin{acknowledgments}
    The author would like to thank Daniela Cadamuro and Henning Bostelmann for suggesting this topic and for many fruitful discussions. Moreover, the author would like to thank Markus Fr\"{o}b for a helpful suggestion on improving numerical convergence.

    The author acknowledges support by the Deutsche Forschungsgemeinschaft (DFG) within the Emmy Noether grant CA1850/1-1 and Grant No 406116891 within the Research Training Group RTG 2522/1. 
\end{acknowledgments}


\begin{thebibliography}{60}%
\makeatletter
\providecommand \@ifxundefined [1]{%
 \@ifx{#1\undefined}
}%
\providecommand \@ifnum [1]{%
 \ifnum #1\expandafter \@firstoftwo
 \else \expandafter \@secondoftwo
 \fi
}%
\providecommand \@ifx [1]{%
 \ifx #1\expandafter \@firstoftwo
 \else \expandafter \@secondoftwo
 \fi
}%
\providecommand \natexlab [1]{#1}%
\providecommand \enquote  [1]{``#1''}%
\providecommand \bibnamefont  [1]{#1}%
\providecommand \bibfnamefont [1]{#1}%
\providecommand \citenamefont [1]{#1}%
\providecommand \href@noop [0]{\@secondoftwo}%
\providecommand \href [0]{\begingroup \@sanitize@url \@href}%
\providecommand \@href[1]{\@@startlink{#1}\@@href}%
\providecommand \@@href[1]{\endgroup#1\@@endlink}%
\providecommand \@sanitize@url [0]{\catcode `\\12\catcode `\$12\catcode
  `\&12\catcode `\#12\catcode `\^12\catcode `\_12\catcode `\%12\relax}%
\providecommand \@@startlink[1]{}%
\providecommand \@@endlink[0]{}%
\providecommand \url  [0]{\begingroup\@sanitize@url \@url }%
\providecommand \@url [1]{\endgroup\@href {#1}{\urlprefix }}%
\providecommand \urlprefix  [0]{URL }%
\providecommand \Eprint [0]{\href }%
\providecommand \doibase [0]{https://doi.org/}%
\providecommand \selectlanguage [0]{\@gobble}%
\providecommand \bibinfo  [0]{\@secondoftwo}%
\providecommand \bibfield  [0]{\@secondoftwo}%
\providecommand \translation [1]{[#1]}%
\providecommand \BibitemOpen [0]{}%
\providecommand \bibitemStop [0]{}%
\providecommand \bibitemNoStop [0]{.\EOS\space}%
\providecommand \EOS [0]{\spacefactor3000\relax}%
\providecommand \BibitemShut  [1]{\csname bibitem#1\endcsname}%
\let\auto@bib@innerbib\@empty
\bibitem [{\citenamefont {Curiel}(2017)}]{Cur17}%
  \BibitemOpen
  \bibfield  {author} {\bibinfo {author} {\bibfnamefont {E.}~\bibnamefont
  {Curiel}},\ }\bibfield  {title} {\bibinfo {title} {A {{Primer}} on {{Energy
  Conditions}}},\ }in\ \href {https://doi.org/10.1007/978-1-4939-3210-8_3}
  {\emph {\bibinfo {booktitle} {Towards a {{Theory}} of {{Spacetime
  Theories}}}}},\ Vol.~\bibinfo {volume} {13},\ \bibinfo {editor} {edited by\
  \bibinfo {editor} {\bibfnamefont {D.}~\bibnamefont {Lehmkuhl}}, \bibinfo
  {editor} {\bibfnamefont {G.}~\bibnamefont {Schiemann}},\ and\ \bibinfo
  {editor} {\bibfnamefont {E.}~\bibnamefont {Scholz}}}\ (\bibinfo  {publisher}
  {{Springer New York}},\ \bibinfo {address} {{New York, NY}},\ \bibinfo {year}
  {2017})\ pp.\ \bibinfo {pages} {43--104}\BibitemShut {NoStop}%
\bibitem [{\citenamefont {Epstein}\ \emph {et~al.}(1965)\citenamefont
  {Epstein}, \citenamefont {Glaser},\ and\ \citenamefont {Jaffe}}]{EGJ65}%
  \BibitemOpen
  \bibfield  {author} {\bibinfo {author} {\bibfnamefont {H.}~\bibnamefont
  {Epstein}}, \bibinfo {author} {\bibfnamefont {V.}~\bibnamefont {Glaser}},\
  and\ \bibinfo {author} {\bibfnamefont {A.}~\bibnamefont {Jaffe}},\ }\bibfield
   {title} {\bibinfo {title} {Nonpositivity of the energy density in quantized
  field theories},\ }\href {https://doi.org/10.1007/BF02749799} {\bibfield
  {journal} {\bibinfo  {journal} {Il Nuovo Cimento (1955-1965)}\ }\textbf
  {\bibinfo {volume} {36}},\ \bibinfo {pages} {1016} (\bibinfo {year}
  {1965})}\BibitemShut {NoStop}%
\bibitem [{\citenamefont {Fewster}(2006)}]{Few06}%
  \BibitemOpen
  \bibfield  {author} {\bibinfo {author} {\bibfnamefont {C.~J.}\ \bibnamefont
  {Fewster}},\ }\bibfield  {title} {\bibinfo {title} {Energy inequalities in
  quantum field theory},\ }in\ \href
  {https://doi.org/10.1142/9789812704016_0056} {\emph {\bibinfo {booktitle}
  {{{XIVth International Congress}} on {{Mathematical Physics}}}}}\ (\bibinfo
  {publisher} {{World Scientific}},\ \bibinfo {address} {{Lisbon, Portugal}},\
  \bibinfo {year} {2006})\ pp.\ \bibinfo {pages} {559--568}\BibitemShut
  {NoStop}%
\bibitem [{\citenamefont {Kontou}\ and\ \citenamefont {Sanders}(2020)}]{KS20}%
  \BibitemOpen
  \bibfield  {author} {\bibinfo {author} {\bibfnamefont {E.-A.}\ \bibnamefont
  {Kontou}}\ and\ \bibinfo {author} {\bibfnamefont {K.}~\bibnamefont
  {Sanders}},\ }\bibfield  {title} {\bibinfo {title} {Energy conditions in
  general relativity and quantum field theory},\ }\href
  {https://doi.org/10.1088/1361-6382/ab8fcf} {\bibfield  {journal} {\bibinfo
  {journal} {Classical and Quantum Gravity}\ }\textbf {\bibinfo {volume}
  {37}},\ \bibinfo {pages} {193001} (\bibinfo {year} {2020})},\ \Eprint
  {https://arxiv.org/abs/2003.01815} {arxiv:2003.01815} \BibitemShut {NoStop}%
\bibitem [{\citenamefont {Ford}(1991)}]{For91}%
  \BibitemOpen
  \bibfield  {author} {\bibinfo {author} {\bibfnamefont {L.~H.}\ \bibnamefont
  {Ford}},\ }\bibfield  {title} {\bibinfo {title} {Constraints on
  negative-energy fluxes},\ }\href {https://doi.org/10.1103/PhysRevD.43.3972}
  {\bibfield  {journal} {\bibinfo  {journal} {Physical Review D}\ }\textbf
  {\bibinfo {volume} {43}},\ \bibinfo {pages} {3972} (\bibinfo {year}
  {1991})}\BibitemShut {NoStop}%
\bibitem [{\citenamefont {Ford}\ and\ \citenamefont {Roman}(1995)}]{FR95}%
  \BibitemOpen
  \bibfield  {author} {\bibinfo {author} {\bibfnamefont {L.~H.}\ \bibnamefont
  {Ford}}\ and\ \bibinfo {author} {\bibfnamefont {T.~A.}\ \bibnamefont
  {Roman}},\ }\bibfield  {title} {\bibinfo {title} {Averaged energy conditions
  and quantum inequalities},\ }\href {https://doi.org/10.1103/PhysRevD.51.4277}
  {\bibfield  {journal} {\bibinfo  {journal} {Physical Review D}\ }\textbf
  {\bibinfo {volume} {51}},\ \bibinfo {pages} {4277} (\bibinfo {year}
  {1995})}\BibitemShut {NoStop}%
\bibitem [{\citenamefont {Ford}\ and\ \citenamefont {Roman}(1997)}]{FR97}%
  \BibitemOpen
  \bibfield  {author} {\bibinfo {author} {\bibfnamefont {L.~H.}\ \bibnamefont
  {Ford}}\ and\ \bibinfo {author} {\bibfnamefont {T.~A.}\ \bibnamefont
  {Roman}},\ }\bibfield  {title} {\bibinfo {title} {Restrictions on negative
  energy density in flat spacetime},\ }\href
  {https://doi.org/10.1103/PhysRevD.55.2082} {\bibfield  {journal} {\bibinfo
  {journal} {Physical Review D}\ }\textbf {\bibinfo {volume} {55}},\ \bibinfo
  {pages} {2082} (\bibinfo {year} {1997})}\BibitemShut {NoStop}%
\bibitem [{\citenamefont {Pfenning}\ and\ \citenamefont {Ford}(1997)}]{PF97}%
  \BibitemOpen
  \bibfield  {author} {\bibinfo {author} {\bibfnamefont {M.~J.}\ \bibnamefont
  {Pfenning}}\ and\ \bibinfo {author} {\bibfnamefont {L.~H.}\ \bibnamefont
  {Ford}},\ }\bibfield  {title} {\bibinfo {title} {Quantum inequalities on the
  energy density in static {{Robertson-Walker}} spacetimes},\ }\href
  {https://doi.org/10.1103/PhysRevD.55.4813} {\bibfield  {journal} {\bibinfo
  {journal} {Physical Review D}\ }\textbf {\bibinfo {volume} {55}},\ \bibinfo
  {pages} {4813} (\bibinfo {year} {1997})}\BibitemShut {NoStop}%
\bibitem [{\citenamefont {Flanagan}(1997)}]{Fla97}%
  \BibitemOpen
  \bibfield  {author} {\bibinfo {author} {\bibfnamefont {E.~E.}\ \bibnamefont
  {Flanagan}},\ }\bibfield  {title} {\bibinfo {title} {Quantum inequalities in
  two-dimensional {{Minkowski}} spacetime},\ }\href
  {https://doi.org/10.1103/PhysRevD.56.4922} {\bibfield  {journal} {\bibinfo
  {journal} {Physical Review D}\ }\textbf {\bibinfo {volume} {56}},\ \bibinfo
  {pages} {4922} (\bibinfo {year} {1997})}\BibitemShut {NoStop}%
\bibitem [{\citenamefont {Fewster}\ and\ \citenamefont {Eveson}(1998)}]{FE98}%
  \BibitemOpen
  \bibfield  {author} {\bibinfo {author} {\bibfnamefont {C.~J.}\ \bibnamefont
  {Fewster}}\ and\ \bibinfo {author} {\bibfnamefont {S.~P.}\ \bibnamefont
  {Eveson}},\ }\bibfield  {title} {\bibinfo {title} {Bounds on negative energy
  densities in flat spacetime},\ }\href
  {https://doi.org/10.1103/PhysRevD.58.084010} {\bibfield  {journal} {\bibinfo
  {journal} {Physical Review D}\ }\textbf {\bibinfo {volume} {58}},\ \bibinfo
  {pages} {084010} (\bibinfo {year} {1998})}\BibitemShut {NoStop}%
\bibitem [{\citenamefont {Ford}\ \emph {et~al.}(1998)\citenamefont {Ford},
  \citenamefont {Pfenning},\ and\ \citenamefont {Roman}}]{FPR98}%
  \BibitemOpen
  \bibfield  {author} {\bibinfo {author} {\bibfnamefont {L.~H.}\ \bibnamefont
  {Ford}}, \bibinfo {author} {\bibfnamefont {M.~J.}\ \bibnamefont {Pfenning}},\
  and\ \bibinfo {author} {\bibfnamefont {T.~A.}\ \bibnamefont {Roman}},\
  }\bibfield  {title} {\bibinfo {title} {Quantum inequalities and singular
  negative energy densities},\ }\href
  {https://doi.org/10.1103/PhysRevD.57.4839} {\bibfield  {journal} {\bibinfo
  {journal} {Physical Review D}\ }\textbf {\bibinfo {volume} {57}},\ \bibinfo
  {pages} {4839} (\bibinfo {year} {1998})}\BibitemShut {NoStop}%
\bibitem [{\citenamefont {Pfenning}\ and\ \citenamefont
  {Ford}(1998{\natexlab{a}})}]{PF98}%
  \BibitemOpen
  \bibfield  {author} {\bibinfo {author} {\bibfnamefont {M.~J.}\ \bibnamefont
  {Pfenning}}\ and\ \bibinfo {author} {\bibfnamefont {L.~H.}\ \bibnamefont
  {Ford}},\ }\href {https://doi.org/10.48550/arXiv.gr-qc/9805037} {\bibinfo
  {title} {Quantum {{Inequality Restrictions}} on {{Negative Energy Densities}}
  in {{Curved Spacetimes}}}} (\bibinfo {year} {1998}{\natexlab{a}}),\ \Eprint
  {https://arxiv.org/abs/gr-qc/9805037} {arxiv:gr-qc/9805037} \BibitemShut
  {NoStop}%
\bibitem [{\citenamefont {Pfenning}\ and\ \citenamefont
  {Ford}(1998{\natexlab{b}})}]{PF98a}%
  \BibitemOpen
  \bibfield  {author} {\bibinfo {author} {\bibfnamefont {M.~J.}\ \bibnamefont
  {Pfenning}}\ and\ \bibinfo {author} {\bibfnamefont {L.~H.}\ \bibnamefont
  {Ford}},\ }\bibfield  {title} {\bibinfo {title} {Scalar field quantum
  inequalities in static spacetimes},\ }\href
  {https://doi.org/10.1103/PhysRevD.57.3489} {\bibfield  {journal} {\bibinfo
  {journal} {Physical Review D}\ }\textbf {\bibinfo {volume} {57}},\ \bibinfo
  {pages} {3489} (\bibinfo {year} {1998}{\natexlab{b}})}\BibitemShut {NoStop}%
\bibitem [{\citenamefont {Vollick}(1998)}]{Vol98}%
  \BibitemOpen
  \bibfield  {author} {\bibinfo {author} {\bibfnamefont {D.~N.}\ \bibnamefont
  {Vollick}},\ }\bibfield  {title} {\bibinfo {title} {Negative energy density
  states for the {{Dirac}} field in flat space-time},\ }\href
  {https://doi.org/10.1103/PhysRevD.57.3484} {\bibfield  {journal} {\bibinfo
  {journal} {Phys. Rev. D}\ }\textbf {\bibinfo {volume} {57}},\ \bibinfo
  {pages} {3484} (\bibinfo {year} {1998})}\BibitemShut {NoStop}%
\bibitem [{\citenamefont {Fewster}\ and\ \citenamefont {Teo}(1999)}]{FT99}%
  \BibitemOpen
  \bibfield  {author} {\bibinfo {author} {\bibfnamefont {C.~J.}\ \bibnamefont
  {Fewster}}\ and\ \bibinfo {author} {\bibfnamefont {E.}~\bibnamefont {Teo}},\
  }\bibfield  {title} {\bibinfo {title} {Bounds on negative energy densities in
  static space-times},\ }\href {https://doi.org/10.1103/PhysRevD.59.104016}
  {\bibfield  {journal} {\bibinfo  {journal} {Physical Review D}\ }\textbf
  {\bibinfo {volume} {59}},\ \bibinfo {pages} {104016} (\bibinfo {year}
  {1999})}\BibitemShut {NoStop}%
\bibitem [{\citenamefont {Fewster}(2000)}]{Few00}%
  \BibitemOpen
  \bibfield  {author} {\bibinfo {author} {\bibfnamefont {C.~J.}\ \bibnamefont
  {Fewster}},\ }\bibfield  {title} {\bibinfo {title} {A general worldline
  quantum inequality},\ }\href {https://doi.org/10.1088/0264-9381/17/9/302}
  {\bibfield  {journal} {\bibinfo  {journal} {Classical and Quantum Gravity}\
  }\textbf {\bibinfo {volume} {17}},\ \bibinfo {pages} {1897} (\bibinfo {year}
  {2000})}\BibitemShut {NoStop}%
\bibitem [{\citenamefont {Vollick}(2000)}]{Vol00}%
  \BibitemOpen
  \bibfield  {author} {\bibinfo {author} {\bibfnamefont {D.~N.}\ \bibnamefont
  {Vollick}},\ }\bibfield  {title} {\bibinfo {title} {Quantum inequalities in
  curved two-dimensional spacetimes},\ }\href
  {https://doi.org/10.1103/PhysRevD.61.084022} {\bibfield  {journal} {\bibinfo
  {journal} {Physical Review D}\ }\textbf {\bibinfo {volume} {61}},\ \bibinfo
  {pages} {084022} (\bibinfo {year} {2000})}\BibitemShut {NoStop}%
\bibitem [{\citenamefont {Fewster}\ and\ \citenamefont {Verch}(2002)}]{FV02}%
  \BibitemOpen
  \bibfield  {author} {\bibinfo {author} {\bibfnamefont {C.~J.}\ \bibnamefont
  {Fewster}}\ and\ \bibinfo {author} {\bibfnamefont {R.}~\bibnamefont
  {Verch}},\ }\bibfield  {title} {\bibinfo {title} {A {{Quantum Weak Energy
  Inequality}} for {{Dirac}} fields in curved spacetime},\ }\href
  {https://doi.org/10.1007/s002200100584} {\bibfield  {journal} {\bibinfo
  {journal} {Communications in Mathematical Physics}\ }\textbf {\bibinfo
  {volume} {225}},\ \bibinfo {pages} {331} (\bibinfo {year} {2002})},\ \Eprint
  {https://arxiv.org/abs/math-ph/0105027} {arxiv:math-ph/0105027} \BibitemShut
  {NoStop}%
\bibitem [{\citenamefont {Pfenning}(2001)}]{Pfe01}%
  \BibitemOpen
  \bibfield  {author} {\bibinfo {author} {\bibfnamefont {M.~J.}\ \bibnamefont
  {Pfenning}},\ }\bibfield  {title} {\bibinfo {title} {Quantum inequalities for
  the electromagnetic field},\ }\href
  {https://doi.org/10.1103/PhysRevD.65.024009} {\bibfield  {journal} {\bibinfo
  {journal} {Physical Review D}\ }\textbf {\bibinfo {volume} {65}},\ \bibinfo
  {pages} {024009} (\bibinfo {year} {2001})}\BibitemShut {NoStop}%
\bibitem [{\citenamefont {Fewster}\ and\ \citenamefont {Mistry}(2003)}]{FM03}%
  \BibitemOpen
  \bibfield  {author} {\bibinfo {author} {\bibfnamefont {C.~J.}\ \bibnamefont
  {Fewster}}\ and\ \bibinfo {author} {\bibfnamefont {B.}~\bibnamefont
  {Mistry}},\ }\bibfield  {title} {\bibinfo {title} {Quantum {{Weak Energy
  Inequalities}} for the {{Dirac}} field in {{Flat Spacetime}}},\ }\href
  {https://doi.org/10.1103/PhysRevD.68.105010} {\bibfield  {journal} {\bibinfo
  {journal} {Physical Review D}\ }\textbf {\bibinfo {volume} {68}},\ \bibinfo
  {pages} {105010} (\bibinfo {year} {2003})},\ \Eprint
  {https://arxiv.org/abs/gr-qc/0307098} {arxiv:gr-qc/0307098} \BibitemShut
  {NoStop}%
\bibitem [{\citenamefont {Graham}\ and\ \citenamefont {Olum}(2003)}]{GO03}%
  \BibitemOpen
  \bibfield  {author} {\bibinfo {author} {\bibfnamefont {N.}~\bibnamefont
  {Graham}}\ and\ \bibinfo {author} {\bibfnamefont {K.~D.}\ \bibnamefont
  {Olum}},\ }\bibfield  {title} {\bibinfo {title} {Negative energy densities in
  quantum field theory with a background potential},\ }\href
  {https://doi.org/10.1103/PhysRevD.67.085014} {\bibfield  {journal} {\bibinfo
  {journal} {Physical Review D}\ }\textbf {\bibinfo {volume} {67}},\ \bibinfo
  {pages} {085014} (\bibinfo {year} {2003})}\BibitemShut {NoStop}%
\bibitem [{\citenamefont {Fewster}(2004)}]{Few04}%
  \BibitemOpen
  \bibfield  {author} {\bibinfo {author} {\bibfnamefont {C.~J.}\ \bibnamefont
  {Fewster}},\ }\bibfield  {title} {\bibinfo {title} {Quantum energy
  inequalities in two dimensions},\ }\href
  {https://doi.org/10.1103/PhysRevD.70.127501} {\bibfield  {journal} {\bibinfo
  {journal} {Physical Review D}\ }\textbf {\bibinfo {volume} {70}},\ \bibinfo
  {pages} {127501} (\bibinfo {year} {2004})}\BibitemShut {NoStop}%
\bibitem [{\citenamefont {Graham}\ \emph {et~al.}(2004)\citenamefont {Graham},
  \citenamefont {Olum},\ and\ \citenamefont {{Schwartz-Perlov}}}]{GOS04}%
  \BibitemOpen
  \bibfield  {author} {\bibinfo {author} {\bibfnamefont {N.}~\bibnamefont
  {Graham}}, \bibinfo {author} {\bibfnamefont {K.~D.}\ \bibnamefont {Olum}},\
  and\ \bibinfo {author} {\bibfnamefont {D.}~\bibnamefont
  {{Schwartz-Perlov}}},\ }\bibfield  {title} {\bibinfo {title} {Energy
  conditions outside a dielectric ball},\ }\href
  {https://doi.org/10.1103/PhysRevD.70.105019} {\bibfield  {journal} {\bibinfo
  {journal} {Physical Review D}\ }\textbf {\bibinfo {volume} {70}},\ \bibinfo
  {pages} {105019} (\bibinfo {year} {2004})},\ \Eprint
  {https://arxiv.org/abs/gr-qc/0407006} {arxiv:gr-qc/0407006} \BibitemShut
  {NoStop}%
\bibitem [{\citenamefont {Fewster}\ and\ \citenamefont
  {Hollands}(2005)}]{FH05}%
  \BibitemOpen
  \bibfield  {author} {\bibinfo {author} {\bibfnamefont {C.~J.}\ \bibnamefont
  {Fewster}}\ and\ \bibinfo {author} {\bibfnamefont {S.}~\bibnamefont
  {Hollands}},\ }\bibfield  {title} {\bibinfo {title} {Quantum energy
  inequalities in two-dimensional conformal field theory},\ }\href
  {https://doi.org/10.1142/s0129055x05002406} {\bibfield  {journal} {\bibinfo
  {journal} {Reviews in Mathematical Physics}\ }\textbf {\bibinfo {volume}
  {17}},\ \bibinfo {pages} {577} (\bibinfo {year} {2005})}\BibitemShut
  {NoStop}%
\bibitem [{\citenamefont {Dawson}(2006)}]{Daw06}%
  \BibitemOpen
  \bibfield  {author} {\bibinfo {author} {\bibfnamefont {S.~P.}\ \bibnamefont
  {Dawson}},\ }\bibfield  {title} {\bibinfo {title} {A quantum weak energy
  inequality for the {{Dirac}} field in two-dimensional flat spacetime},\
  }\href {https://doi.org/10.1088/0264-9381/23/1/014} {\bibfield  {journal}
  {\bibinfo  {journal} {Classical and Quantum Gravity}\ }\textbf {\bibinfo
  {volume} {23}},\ \bibinfo {pages} {287} (\bibinfo {year} {2006})}\BibitemShut
  {NoStop}%
\bibitem [{\citenamefont {Dawson}\ and\ \citenamefont {Fewster}(2006)}]{DF06}%
  \BibitemOpen
  \bibfield  {author} {\bibinfo {author} {\bibfnamefont {S.~P.}\ \bibnamefont
  {Dawson}}\ and\ \bibinfo {author} {\bibfnamefont {C.~J.}\ \bibnamefont
  {Fewster}},\ }\bibfield  {title} {\bibinfo {title} {An explicit quantum weak
  energy inequality for {{Dirac}} fields in curved spacetimes},\ }\href
  {https://doi.org/10.1088/0264-9381/23/23/005} {\bibfield  {journal} {\bibinfo
   {journal} {Classical and Quantum Gravity}\ }\textbf {\bibinfo {volume}
  {23}},\ \bibinfo {pages} {6659} (\bibinfo {year} {2006})}\BibitemShut
  {NoStop}%
\bibitem [{\citenamefont {Shu}\ \emph {et~al.}(2006)\citenamefont {Shu},
  \citenamefont {Yu}, \citenamefont {Li}, \citenamefont {Wu},\ and\
  \citenamefont {Ren}}]{SYL+06}%
  \BibitemOpen
  \bibfield  {author} {\bibinfo {author} {\bibfnamefont {W.-X.}\ \bibnamefont
  {Shu}}, \bibinfo {author} {\bibfnamefont {H.-W.}\ \bibnamefont {Yu}},
  \bibinfo {author} {\bibfnamefont {F.}~\bibnamefont {Li}}, \bibinfo {author}
  {\bibfnamefont {P.-X.}\ \bibnamefont {Wu}},\ and\ \bibinfo {author}
  {\bibfnamefont {Z.-Z.}\ \bibnamefont {Ren}},\ }\bibfield  {title} {\bibinfo
  {title} {Quantum inequality bounds for free {{Rarita-Schwinger}} field in
  flat spacetime},\ }\href {https://doi.org/10.1088/0253-6102/46/1/021}
  {\bibfield  {journal} {\bibinfo  {journal} {Commun. Theor. Phys.}\ }\textbf
  {\bibinfo {volume} {46}},\ \bibinfo {pages} {87} (\bibinfo {year}
  {2006})}\BibitemShut {NoStop}%
\bibitem [{\citenamefont {Fewster}\ and\ \citenamefont
  {Osterbrink}(2008)}]{FO08}%
  \BibitemOpen
  \bibfield  {author} {\bibinfo {author} {\bibfnamefont {C.~J.}\ \bibnamefont
  {Fewster}}\ and\ \bibinfo {author} {\bibfnamefont {L.~W.}\ \bibnamefont
  {Osterbrink}},\ }\bibfield  {title} {\bibinfo {title} {Quantum {{Energy
  Inequalities}} for the {{Non-Minimally Coupled Scalar Field}}},\ }\href
  {https://doi.org/10.1088/1751-8113/41/2/025402} {\bibfield  {journal}
  {\bibinfo  {journal} {Journal of Physics A: Mathematical and Theoretical}\
  }\textbf {\bibinfo {volume} {41}},\ \bibinfo {pages} {025402} (\bibinfo
  {year} {2008})},\ \Eprint {https://arxiv.org/abs/0708.2450} {arxiv:0708.2450
  [gr-qc]} \BibitemShut {NoStop}%
\bibitem [{\citenamefont {Fewster}\ and\ \citenamefont {Smith}(2008)}]{FS08}%
  \BibitemOpen
  \bibfield  {author} {\bibinfo {author} {\bibfnamefont {C.~J.}\ \bibnamefont
  {Fewster}}\ and\ \bibinfo {author} {\bibfnamefont {C.~J.}\ \bibnamefont
  {Smith}},\ }\bibfield  {title} {\bibinfo {title} {Absolute {{Quantum Energy
  Inequalities}} in {{Curved Spacetime}}},\ }\href
  {https://doi.org/10.1007/s00023-008-0361-0} {\bibfield  {journal} {\bibinfo
  {journal} {Annales Henri Poincar{\'e}}\ }\textbf {\bibinfo {volume} {9}},\
  \bibinfo {pages} {425} (\bibinfo {year} {2008})}\BibitemShut {NoStop}%
\bibitem [{\citenamefont {Schlemmer}\ and\ \citenamefont {Verch}(2008)}]{SV08}%
  \BibitemOpen
  \bibfield  {author} {\bibinfo {author} {\bibfnamefont {J.}~\bibnamefont
  {Schlemmer}}\ and\ \bibinfo {author} {\bibfnamefont {R.}~\bibnamefont
  {Verch}},\ }\bibfield  {title} {\bibinfo {title} {Local {{Thermal Equilibrium
  States}} and {{Quantum Energy Inequalities}}},\ }\href
  {https://doi.org/10.1007/s00023-008-0380-x} {\bibfield  {journal} {\bibinfo
  {journal} {Annales Henri Poincar{\'e}}\ }\textbf {\bibinfo {volume} {9}},\
  \bibinfo {pages} {945} (\bibinfo {year} {2008})}\BibitemShut {NoStop}%
\bibitem [{\citenamefont {Klinkhammer}(1991)}]{Kli91}%
  \BibitemOpen
  \bibfield  {author} {\bibinfo {author} {\bibfnamefont {G.}~\bibnamefont
  {Klinkhammer}},\ }\bibfield  {title} {\bibinfo {title} {Averaged energy
  conditions for free scalar fields in flat spacetime},\ }\href
  {https://doi.org/10.1103/PhysRevD.43.2542} {\bibfield  {journal} {\bibinfo
  {journal} {Physical Review D}\ }\textbf {\bibinfo {volume} {43}},\ \bibinfo
  {pages} {2542} (\bibinfo {year} {1991})}\BibitemShut {NoStop}%
\bibitem [{\citenamefont {Folacci}(1992)}]{Fol92a}%
  \BibitemOpen
  \bibfield  {author} {\bibinfo {author} {\bibfnamefont {A.}~\bibnamefont
  {Folacci}},\ }\bibfield  {title} {\bibinfo {title} {Averaged-null-energy
  condition for electromagnetism in {{Minkowski}} spacetime},\ }\href
  {https://doi.org/10.1103/PhysRevD.46.2726} {\bibfield  {journal} {\bibinfo
  {journal} {Physical Review D}\ }\textbf {\bibinfo {volume} {46}},\ \bibinfo
  {pages} {2726} (\bibinfo {year} {1992})}\BibitemShut {NoStop}%
\bibitem [{\citenamefont {Yurtsever}(1995)}]{Yur95}%
  \BibitemOpen
  \bibfield  {author} {\bibinfo {author} {\bibfnamefont {U.}~\bibnamefont
  {Yurtsever}},\ }\bibfield  {title} {\bibinfo {title} {Averaged null energy
  condition and difference inequalities in quantum field theory},\ }\href
  {https://doi.org/10.1103/PhysRevD.51.5797} {\bibfield  {journal} {\bibinfo
  {journal} {Physical Review D}\ }\textbf {\bibinfo {volume} {51}},\ \bibinfo
  {pages} {5797} (\bibinfo {year} {1995})}\BibitemShut {NoStop}%
\bibitem [{\citenamefont {Flanagan}\ and\ \citenamefont {Wald}(1996)}]{FW96}%
  \BibitemOpen
  \bibfield  {author} {\bibinfo {author} {\bibfnamefont {E.~E.}\ \bibnamefont
  {Flanagan}}\ and\ \bibinfo {author} {\bibfnamefont {R.~M.}\ \bibnamefont
  {Wald}},\ }\bibfield  {title} {\bibinfo {title} {Does back reaction enforce
  the averaged null energy condition in semiclassical gravity?},\ }\href
  {https://doi.org/10.1103/PhysRevD.54.6233} {\bibfield  {journal} {\bibinfo
  {journal} {Physical Review D}\ }\textbf {\bibinfo {volume} {54}},\ \bibinfo
  {pages} {6233} (\bibinfo {year} {1996})}\BibitemShut {NoStop}%
\bibitem [{\citenamefont {Ford}\ and\ \citenamefont {Roman}(1996)}]{FR96}%
  \BibitemOpen
  \bibfield  {author} {\bibinfo {author} {\bibfnamefont {L.~H.}\ \bibnamefont
  {Ford}}\ and\ \bibinfo {author} {\bibfnamefont {T.~A.}\ \bibnamefont
  {Roman}},\ }\bibfield  {title} {\bibinfo {title} {Quantum {{Field Theory
  Constrains Traversable Wormhole Geometries}}},\ }\href
  {https://doi.org/10.1103/PhysRevD.53.5496} {\bibfield  {journal} {\bibinfo
  {journal} {Physical Review D}\ }\textbf {\bibinfo {volume} {53}},\ \bibinfo
  {pages} {5496} (\bibinfo {year} {1996})},\ \Eprint
  {https://arxiv.org/abs/gr-qc/9510071} {arxiv:gr-qc/9510071} \BibitemShut
  {NoStop}%
\bibitem [{\citenamefont {Fewster}\ and\ \citenamefont {Roman}(2003)}]{FR03}%
  \BibitemOpen
  \bibfield  {author} {\bibinfo {author} {\bibfnamefont {C.~J.}\ \bibnamefont
  {Fewster}}\ and\ \bibinfo {author} {\bibfnamefont {T.~A.}\ \bibnamefont
  {Roman}},\ }\bibfield  {title} {\bibinfo {title} {Null energy conditions in
  quantum field theory},\ }\href {https://doi.org/10.1103/PhysRevD.67.044003}
  {\bibfield  {journal} {\bibinfo  {journal} {Physical Review D}\ }\textbf
  {\bibinfo {volume} {67}},\ \bibinfo {pages} {044003} (\bibinfo {year}
  {2003})}\BibitemShut {NoStop}%
\bibitem [{\citenamefont {Wald}\ and\ \citenamefont {Yurtsever}(1991)}]{WY91}%
  \BibitemOpen
  \bibfield  {author} {\bibinfo {author} {\bibfnamefont {R.}~\bibnamefont
  {Wald}}\ and\ \bibinfo {author} {\bibfnamefont {U.}~\bibnamefont
  {Yurtsever}},\ }\bibfield  {title} {\bibinfo {title} {General proof of the
  averaged null energy condition for a massless scalar field in two-dimensional
  curved spacetime},\ }\href {https://doi.org/10.1103/PhysRevD.44.403}
  {\bibfield  {journal} {\bibinfo  {journal} {Physical Review D}\ }\textbf
  {\bibinfo {volume} {44}},\ \bibinfo {pages} {403} (\bibinfo {year}
  {1991})}\BibitemShut {NoStop}%
\bibitem [{\citenamefont {Verch}(2000)}]{Ver00}%
  \BibitemOpen
  \bibfield  {author} {\bibinfo {author} {\bibfnamefont {R.}~\bibnamefont
  {Verch}},\ }\bibfield  {title} {\bibinfo {title} {The averaged null energy
  condition for general quantum field theories in two dimensions},\ }\href
  {https://doi.org/10.1063/1.533130} {\bibfield  {journal} {\bibinfo  {journal}
  {Journal of Mathematical Physics}\ }\textbf {\bibinfo {volume} {41}},\
  \bibinfo {pages} {206} (\bibinfo {year} {2000})}\BibitemShut {NoStop}%
\bibitem [{\citenamefont {Bostelmann}\ and\ \citenamefont
  {Fewster}(2009)}]{BF09}%
  \BibitemOpen
  \bibfield  {author} {\bibinfo {author} {\bibfnamefont {H.}~\bibnamefont
  {Bostelmann}}\ and\ \bibinfo {author} {\bibfnamefont {C.~J.}\ \bibnamefont
  {Fewster}},\ }\bibfield  {title} {\bibinfo {title} {Quantum {{Inequalities}}
  from {{Operator Product Expansions}}},\ }\href
  {https://doi.org/10.1007/s00220-009-0853-x} {\bibfield  {journal} {\bibinfo
  {journal} {Communications in Mathematical Physics}\ }\textbf {\bibinfo
  {volume} {292}},\ \bibinfo {pages} {761} (\bibinfo {year}
  {2009})}\BibitemShut {NoStop}%
\bibitem [{\citenamefont {Much}\ \emph {et~al.}(2022)\citenamefont {Much},
  \citenamefont {Passegger},\ and\ \citenamefont {Verch}}]{MPV22}%
  \BibitemOpen
  \bibfield  {author} {\bibinfo {author} {\bibfnamefont {A.}~\bibnamefont
  {Much}}, \bibinfo {author} {\bibfnamefont {A.~G.}\ \bibnamefont
  {Passegger}},\ and\ \bibinfo {author} {\bibfnamefont {R.}~\bibnamefont
  {Verch}},\ }\href {https://doi.org/10.48550/arXiv.2210.01145} {\bibinfo
  {title} {An approximate local modular quantum energy inequality in general
  quantum field theory}} (\bibinfo {year} {2022}),\ \Eprint
  {https://arxiv.org/abs/2210.01145} {arxiv:2210.01145 [gr-qc, physics:hep-th,
  physics:math-ph]} \BibitemShut {NoStop}%
\bibitem [{\citenamefont {Bostelmann}\ \emph {et~al.}(2013)\citenamefont
  {Bostelmann}, \citenamefont {Cadamuro},\ and\ \citenamefont
  {Fewster}}]{BCF13}%
  \BibitemOpen
  \bibfield  {author} {\bibinfo {author} {\bibfnamefont {H.}~\bibnamefont
  {Bostelmann}}, \bibinfo {author} {\bibfnamefont {D.}~\bibnamefont
  {Cadamuro}},\ and\ \bibinfo {author} {\bibfnamefont {C.~J.}\ \bibnamefont
  {Fewster}},\ }\bibfield  {title} {\bibinfo {title} {Quantum energy inequality
  for the massive {{Ising}} model},\ }\href
  {https://doi.org/10.1103/PhysRevD.88.025019} {\bibfield  {journal} {\bibinfo
  {journal} {Physical Review D}\ }\textbf {\bibinfo {volume} {88}},\ \bibinfo
  {pages} {025019} (\bibinfo {year} {2013})}\BibitemShut {NoStop}%
\bibitem [{\citenamefont {Fr{\"o}b}\ and\ \citenamefont
  {Cadamuro}(2022)}]{FC22}%
  \BibitemOpen
  \bibfield  {author} {\bibinfo {author} {\bibfnamefont {M.~B.}\ \bibnamefont
  {Fr{\"o}b}}\ and\ \bibinfo {author} {\bibfnamefont {D.}~\bibnamefont
  {Cadamuro}},\ }\href {https://doi.org/10.48550/arXiv.2212.07377} {\bibinfo
  {title} {A quantum energy inequality in the {{Sine--Gordon}} model}}
  (\bibinfo {year} {2022}),\ \Eprint {https://arxiv.org/abs/2212.07377}
  {arxiv:2212.07377 [hep-th, physics:math-ph]} \BibitemShut {NoStop}%
\bibitem [{\citenamefont {Bostelmann}\ and\ \citenamefont
  {Cadamuro}(2016)}]{BC16}%
  \BibitemOpen
  \bibfield  {author} {\bibinfo {author} {\bibfnamefont {H.}~\bibnamefont
  {Bostelmann}}\ and\ \bibinfo {author} {\bibfnamefont {D.}~\bibnamefont
  {Cadamuro}},\ }\bibfield  {title} {\bibinfo {title} {Negative energy
  densities in integrable quantum field theories at one-particle level},\
  }\href {https://doi.org/10.1103/PhysRevD.93.065001} {\bibfield  {journal}
  {\bibinfo  {journal} {Physical Review D}\ }\textbf {\bibinfo {volume} {93}},\
  \bibinfo {pages} {065001} (\bibinfo {year} {2016})}\BibitemShut {NoStop}%
\bibitem [{\citenamefont {Bostelmann}\ \emph {et~al.}(2023)\citenamefont
  {Bostelmann}, \citenamefont {Cadamuro},\ and\ \citenamefont
  {Mandrysch}}]{BCM23}%
  \BibitemOpen
  \bibfield  {author} {\bibinfo {author} {\bibfnamefont {H.}~\bibnamefont
  {Bostelmann}}, \bibinfo {author} {\bibfnamefont {D.}~\bibnamefont
  {Cadamuro}},\ and\ \bibinfo {author} {\bibfnamefont {J.}~\bibnamefont
  {Mandrysch}},\ }\href {https://doi.org/10.48550/arXiv.2302.00063} {\bibinfo
  {title} {Quantum energy inequalities in integrable models with several
  particle species and bound states}} (\bibinfo {year} {2023}),\ \Eprint
  {https://arxiv.org/abs/2302.00063} {arxiv:2302.00063 [hep-th,
  physics:math-ph]} \BibitemShut {NoStop}%
\bibitem [{\citenamefont {Bostelmann}\ and\ \citenamefont
  {Cadamuro}(2015)}]{BC15}%
  \BibitemOpen
  \bibfield  {author} {\bibinfo {author} {\bibfnamefont {H.}~\bibnamefont
  {Bostelmann}}\ and\ \bibinfo {author} {\bibfnamefont {D.}~\bibnamefont
  {Cadamuro}},\ }\bibfield  {title} {\bibinfo {title} {Characterization of
  {{Local Observables}} in {{Integrable Quantum Field Theories}}},\ }\href
  {https://doi.org/10.1007/s00220-015-2294-z} {\bibfield  {journal} {\bibinfo
  {journal} {Communications in Mathematical Physics}\ }\textbf {\bibinfo
  {volume} {337}},\ \bibinfo {pages} {1199} (\bibinfo {year}
  {2015})}\BibitemShut {NoStop}%
\bibitem [{\citenamefont {Lechner}\ and\ \citenamefont
  {Sch{\"u}tzenhofer}(2014)}]{LS14}%
  \BibitemOpen
  \bibfield  {author} {\bibinfo {author} {\bibfnamefont {G.}~\bibnamefont
  {Lechner}}\ and\ \bibinfo {author} {\bibfnamefont {C.}~\bibnamefont
  {Sch{\"u}tzenhofer}},\ }\bibfield  {title} {\bibinfo {title} {Towards an
  {{Operator-Algebraic Construction}} of {{Integrable Global Gauge
  Theories}}},\ }\href {https://doi.org/10.1007/s00023-013-0260-x} {\bibfield
  {journal} {\bibinfo  {journal} {Annales Henri Poincar{\'e}}\ }\textbf
  {\bibinfo {volume} {15}},\ \bibinfo {pages} {645} (\bibinfo {year}
  {2014})}\BibitemShut {NoStop}%
\bibitem [{\citenamefont {Alazzawi}\ and\ \citenamefont
  {Lechner}(2017)}]{AL17}%
  \BibitemOpen
  \bibfield  {author} {\bibinfo {author} {\bibfnamefont {S.}~\bibnamefont
  {Alazzawi}}\ and\ \bibinfo {author} {\bibfnamefont {G.}~\bibnamefont
  {Lechner}},\ }\bibfield  {title} {\bibinfo {title} {Inverse {{Scattering}}
  and {{Local Observable Algebras}} in {{Integrable Quantum Field Theories}}},\
  }\href {https://doi.org/10.1007/s00220-017-2891-0} {\bibfield  {journal}
  {\bibinfo  {journal} {Communications in Mathematical Physics}\ }\textbf
  {\bibinfo {volume} {354}},\ \bibinfo {pages} {913} (\bibinfo {year}
  {2017})}\BibitemShut {NoStop}%
\bibitem [{\citenamefont {{Mandrysch, Jan}}(2023)}]{Man23}%
  \BibitemOpen
  \bibfield  {author} {\bibinfo {author} {\bibnamefont {{Mandrysch, Jan}}},\
  }\emph {\bibinfo {title} {Energy Inequalities in Integrable Quantum Field
  Theory}},\ \href@noop {} {\bibinfo {type} {Doctoral {{Thesis}}}},\ \bibinfo
  {school} {Universit{\"a}t Leipzig}, \bibinfo {address} {{Leipzig}} (\bibinfo
  {year} {2023})\BibitemShut {NoStop}%
\bibitem [{\citenamefont {Liguori}\ and\ \citenamefont
  {Mintchev}(1995)}]{LM95}%
  \BibitemOpen
  \bibfield  {author} {\bibinfo {author} {\bibfnamefont {A.}~\bibnamefont
  {Liguori}}\ and\ \bibinfo {author} {\bibfnamefont {M.}~\bibnamefont
  {Mintchev}},\ }\bibfield  {title} {\bibinfo {title} {Fock {{Representations}}
  of {{Quantum Fields}} with {{Generalized Statistic}}},\ }\href
  {https://doi.org/10.1007/BF02099316} {\bibfield  {journal} {\bibinfo
  {journal} {Communications in Mathematical Physics}\ }\textbf {\bibinfo
  {volume} {169}},\ \bibinfo {pages} {635} (\bibinfo {year} {1995})},\ \Eprint
  {https://arxiv.org/abs/hep-th/9403039} {arxiv:hep-th/9403039} \BibitemShut
  {NoStop}%
\bibitem [{\citenamefont {Karowski}\ and\ \citenamefont {Weisz}(1978)}]{KW78}%
  \BibitemOpen
  \bibfield  {author} {\bibinfo {author} {\bibfnamefont {M.}~\bibnamefont
  {Karowski}}\ and\ \bibinfo {author} {\bibfnamefont {P.}~\bibnamefont
  {Weisz}},\ }\bibfield  {title} {\bibinfo {title} {Exact form factors in (1 +
  1)-dimensional field theoretic models with soliton behaviour},\ }\href
  {https://doi.org/10.1016/0550-3213(78)90362-0} {\bibfield  {journal}
  {\bibinfo  {journal} {Nuclear Physics B}\ }\textbf {\bibinfo {volume}
  {139}},\ \bibinfo {pages} {455} (\bibinfo {year} {1978})}\BibitemShut
  {NoStop}%
\bibitem [{\citenamefont {Babujian}\ \emph {et~al.}(2008)\citenamefont
  {Babujian}, \citenamefont {Foerster},\ and\ \citenamefont
  {Karowski}}]{BFK08}%
  \BibitemOpen
  \bibfield  {author} {\bibinfo {author} {\bibfnamefont {H.~M.}\ \bibnamefont
  {Babujian}}, \bibinfo {author} {\bibfnamefont {A.}~\bibnamefont {Foerster}},\
  and\ \bibinfo {author} {\bibfnamefont {M.}~\bibnamefont {Karowski}},\
  }\bibfield  {title} {\bibinfo {title} {The {{Form Factor Program}}: {{A
  Review}} and {{New Results}}, the {{Nested Su}}(n) {{Off-Shell Bethe Ansatz}}
  and the 1/{{N Expansion}}},\ }\href
  {https://doi.org/10.1007/s11232-008-0042-7} {\bibfield  {journal} {\bibinfo
  {journal} {Theoretical and Mathematical Physics}\ }\textbf {\bibinfo {volume}
  {155}},\ \bibinfo {pages} {512} (\bibinfo {year} {2008})}\BibitemShut
  {NoStop}%
\bibitem [{\citenamefont {Bostelmann}\ and\ \citenamefont
  {Cadamuro}(2013)}]{BC13}%
  \BibitemOpen
  \bibfield  {author} {\bibinfo {author} {\bibfnamefont {H.}~\bibnamefont
  {Bostelmann}}\ and\ \bibinfo {author} {\bibfnamefont {D.}~\bibnamefont
  {Cadamuro}},\ }\bibfield  {title} {\bibinfo {title} {An operator expansion
  for integrable quantum field theories},\ }\href
  {https://doi.org/10.1088/1751-8113/46/9/095401} {\bibfield  {journal}
  {\bibinfo  {journal} {Journal of Physics A: Mathematical and Theoretical}\
  }\textbf {\bibinfo {volume} {46}},\ \bibinfo {pages} {095401} (\bibinfo
  {year} {2013})},\ \Eprint {https://arxiv.org/abs/1208.4763} {arxiv:1208.4763}
  \BibitemShut {NoStop}%
\bibitem [{\citenamefont {Smirnov}(1992)}]{Smi92}%
  \BibitemOpen
  \bibfield  {author} {\bibinfo {author} {\bibfnamefont {F.~A.}\ \bibnamefont
  {Smirnov}},\ }\href@noop {} {\emph {\bibinfo {title} {Form {{Factors In
  Completely Integrable Models Of Quantum Field Theory}}}}}\ (\bibinfo
  {publisher} {{World Scientific}},\ \bibinfo {year} {1992})\BibitemShut
  {NoStop}%
\bibitem [{\citenamefont {Fring}\ \emph {et~al.}(1993)\citenamefont {Fring},
  \citenamefont {Mussardo},\ and\ \citenamefont {Simonetti}}]{FMS93}%
  \BibitemOpen
  \bibfield  {author} {\bibinfo {author} {\bibfnamefont {A.}~\bibnamefont
  {Fring}}, \bibinfo {author} {\bibfnamefont {G.}~\bibnamefont {Mussardo}},\
  and\ \bibinfo {author} {\bibfnamefont {P.}~\bibnamefont {Simonetti}},\
  }\bibfield  {title} {\bibinfo {title} {Form factors for integrable lagrangian
  field theories, the sinh-{{Gordon}} model},\ }\href
  {https://doi.org/10.1016/0550-3213(93)90252-k} {\bibfield  {journal}
  {\bibinfo  {journal} {Nuclear Physics B}\ }\textbf {\bibinfo {volume}
  {393}},\ \bibinfo {pages} {413} (\bibinfo {year} {1993})}\BibitemShut
  {NoStop}%
\bibitem [{\citenamefont {{Castro-Alvaredo}}\ and\ \citenamefont
  {Fring}(2001)}]{CF01}%
  \BibitemOpen
  \bibfield  {author} {\bibinfo {author} {\bibfnamefont {O.}~\bibnamefont
  {{Castro-Alvaredo}}}\ and\ \bibinfo {author} {\bibfnamefont {A.}~\bibnamefont
  {Fring}},\ }\bibfield  {title} {\bibinfo {title} {Form factors from free
  fermionic {{Fock}} fields, the {{Federbush}} model},\ }\href
  {https://doi.org/10.1016/S0550-3213(01)00462-X} {\bibfield  {journal}
  {\bibinfo  {journal} {Nuclear Physics B}\ }\textbf {\bibinfo {volume}
  {618}},\ \bibinfo {pages} {437} (\bibinfo {year} {2001})}\BibitemShut
  {NoStop}%
\bibitem [{\citenamefont {Babujian}\ \emph {et~al.}(2006)\citenamefont
  {Babujian}, \citenamefont {Foerster},\ and\ \citenamefont
  {Karowski}}]{BFK06}%
  \BibitemOpen
  \bibfield  {author} {\bibinfo {author} {\bibfnamefont {H.}~\bibnamefont
  {Babujian}}, \bibinfo {author} {\bibfnamefont {A.}~\bibnamefont {Foerster}},\
  and\ \bibinfo {author} {\bibfnamefont {M.}~\bibnamefont {Karowski}},\
  }\bibfield  {title} {\bibinfo {title} {Exact form factors in integrable
  quantum field theories: The scaling {{Z}}({{N}})-{{Ising}} model},\ }\href
  {https://doi.org/10.1016/j.nuclphysb.2005.12.001} {\bibfield  {journal}
  {\bibinfo  {journal} {Nuclear Physics B}\ }\textbf {\bibinfo {volume}
  {736}},\ \bibinfo {pages} {169} (\bibinfo {year} {2006})},\ \Eprint
  {https://arxiv.org/abs/hep-th/0510062} {arxiv:hep-th/0510062} \BibitemShut
  {NoStop}%
\bibitem [{\citenamefont {Arinshtein}\ \emph {et~al.}(1979)\citenamefont
  {Arinshtein}, \citenamefont {Fateyev},\ and\ \citenamefont
  {Zamolodchikov}}]{AFZ79}%
  \BibitemOpen
  \bibfield  {author} {\bibinfo {author} {\bibfnamefont {A.~E.}\ \bibnamefont
  {Arinshtein}}, \bibinfo {author} {\bibfnamefont {V.~A.}\ \bibnamefont
  {Fateyev}},\ and\ \bibinfo {author} {\bibfnamefont {A.~B.}\ \bibnamefont
  {Zamolodchikov}},\ }\bibfield  {title} {\bibinfo {title} {Quantum
  {{S-matrix}} of the (1 + 1)-dimensional {{Todd}} chain},\ }\href
  {https://doi.org/10.1016/0370-2693(79)90561-6} {\bibfield  {journal}
  {\bibinfo  {journal} {Physics Letters B}\ }\textbf {\bibinfo {volume} {87}},\
  \bibinfo {pages} {389} (\bibinfo {year} {1979})}\BibitemShut {NoStop}%
\bibitem [{\citenamefont {Braden}\ and\ \citenamefont {Sasaki}(1991)}]{BS91}%
  \BibitemOpen
  \bibfield  {author} {\bibinfo {author} {\bibfnamefont {H.}~\bibnamefont
  {Braden}}\ and\ \bibinfo {author} {\bibfnamefont {R.}~\bibnamefont
  {Sasaki}},\ }\bibfield  {title} {\bibinfo {title} {The {{S-matrix}} coupling
  dependence for a, d and e affine {{Toda}} field theory},\ }\href
  {https://doi.org/10.1016/0370-2693(91)90777-N} {\bibfield  {journal}
  {\bibinfo  {journal} {Physics Letters B}\ }\textbf {\bibinfo {volume}
  {255}},\ \bibinfo {pages} {343} (\bibinfo {year} {1991})}\BibitemShut
  {NoStop}%
\bibitem [{\citenamefont {Klassen}\ and\ \citenamefont {Melzer}(1993)}]{KM93}%
  \BibitemOpen
  \bibfield  {author} {\bibinfo {author} {\bibfnamefont {T.~R.}\ \bibnamefont
  {Klassen}}\ and\ \bibinfo {author} {\bibfnamefont {E.}~\bibnamefont
  {Melzer}},\ }\bibfield  {title} {\bibinfo {title} {Sine-{{Gordon}} =/=
  {{Massive Thirring}}, and {{Related Heresies}}},\ }\href
  {https://doi.org/10.1142/S0217751X93001703} {\bibfield  {journal} {\bibinfo
  {journal} {International Journal of Modern Physics A}\ }\textbf {\bibinfo
  {volume} {08}},\ \bibinfo {pages} {4131} (\bibinfo {year} {1993})},\ \Eprint
  {https://arxiv.org/abs/hep-th/9206114} {arxiv:hep-th/9206114} \BibitemShut
  {NoStop}%
\bibitem [{\citenamefont {Mussardo}\ and\ \citenamefont
  {Simonetti}(1994)}]{MS94}%
  \BibitemOpen
  \bibfield  {author} {\bibinfo {author} {\bibfnamefont {G.}~\bibnamefont
  {Mussardo}}\ and\ \bibinfo {author} {\bibfnamefont {P.}~\bibnamefont
  {Simonetti}},\ }\bibfield  {title} {\bibinfo {title} {Stress-{{Energy
  Tensor}} and {{Ultraviolet Behaviour}} in {{Massive Integrable Quantum Field
  Theories}}},\ }\href {https://doi.org/10.1142/S0217751X94001308} {\bibfield
  {journal} {\bibinfo  {journal} {International Journal of Modern Physics A}\
  }\textbf {\bibinfo {volume} {09}},\ \bibinfo {pages} {3307} (\bibinfo {year}
  {1994})},\ \Eprint {https://arxiv.org/abs/hep-th/9308057}
  {arxiv:hep-th/9308057} \BibitemShut {NoStop}%
\end{thebibliography}
%

\end{document}